\documentclass[12pt]{iopart}
\usepackage{iopams}  

\usepackage{bm,bbm}
\usepackage{verbatim}
\expandafter\let\csname equation*\endcsname\relax

\expandafter\let\csname endequation*\endcsname\relax
\usepackage{amsmath}

\usepackage[capitalise]{cleveref}
\usepackage{dsfont}
\usepackage{lipsum}
\usepackage{xcolor}
\usepackage{soul}
\usepackage{caption}
\usepackage{subcaption}
\usepackage{graphicx}
\newcommand{\Log}{\mathrm{Log}}
\usepackage{cite}

\begin{document}

\title{Top eigenpair \color{black} statistics of diluted Wishart matrices}

\author[1]{Barak Budnick$^1$, Preben Forer$^2$, Pierpaolo Vivo$^2$, Sabrina Aufiero$^3$, Silvia Bartolucci$^3$, Fabio Caccioli$^{3,4,5}$}

\address{$^1$Racah Institute of Physics, The Hebrew University, Jerusalem 9190401, Israel}
\address{$^2$Department of Mathematics, King’s College London, Strand, London, WC2R 2LS, United Kingdom}
\address{$^3$Department of Computer Science, University College London, 66-72 Gower Street, London, WC1E 6EA, United Kingdom}
\address{$^4$Systemic Risk Centre, London School of Economics and Political Sciences, London, WC2A 2AE, United Kingdom}
\address{$^5$London Mathematical Laboratory, 8 Margravine Gardens, London, WC 8RH, United Kingdom}

\vspace{10pt}
\begin{indented}
\item[]December 2024
\end{indented}

\begin{abstract}
Using the replica method, we compute the statistics of the top eigenpair \color{black} of diluted covariance matrices of the form $\bm J = \bm X^T \bm X$, where $\bm X$ is a $N\times M$ sparse data matrix, in the limit of large $N,M$ with fixed ratio and a bounded number of nonzero entries. \color{black} We allow for random non-zero weights, provided they lead to an isolated largest eigenvalue. By formulating the problem as the optimisation of a quadratic Hamiltonian constrained to the $N$-sphere at low temperatures, we derive a set of recursive distributional equations for auxiliary probability density functions, which can be efficiently solved using a population dynamics algorithm. The average largest eigenvalue is identified with a Lagrange parameter that governs the convergence of the algorithm, and the resulting stable populations are then used to evaluate the density of the top eigenvector's components. \color{black} We find excellent agreement between our analytical results and numerical results obtained from direct diagonalisation.
\end{abstract}

\section{Introduction}
\label{introduction}

In recent decades, we have witnessed an unprecedented surge in the amount of information available for processing and forecasting, marking the emergence of the Big Data era. Contemporary data analysis challenges frequently involve processing datasets with numerous variables and observations. This high-dimensional nature of data is particularly evident in fields such as climate studies, genetics, biomedical imaging, and economics \cite{Fan2014}.

Consider a scenario where one conducts $N$ measurements of $M$ variables that characterise a system. These variables might represent, for instance, assets in a stock market or a collection of climate observables, with measurements taken simultaneously at $N$ different time points. The collected data can be organised into an $N \times M$ matrix $\bm X$, where element $X_{ij}$ represents the $i$-th measurement of the $j$-th variable. From this, we construct the $M \times M$ sample covariance matrix $\bm J = \bm X^T \bm X$, which encodes all possible correlations among the variables. This covariance matrix plays a fundamental role in multivariate statistical analysis, finding applications in dimensional reduction and classification procedures, such as Principal Component Analysis \cite{bookPCA} and linear discriminant analysis \cite{bookLDA}.

A reasonable assumption for many natural phenomena is that each variable exhibits significant correlation with only a limited subset of other variables, resulting in sparse covariance matrices characterised by numerous entries that are either very small or zero. This sparsity is particularly relevant in inferring causal influences among system components from empirical covariance matrices. Notable examples include the experimental reconstruction of interactions in biological systems, such as cellular signalling networks \cite{Sachs2005}, gene regulatory networks \cite{Butte2000,Witten2009}, and ecological association networks \cite{Deng2012,Kurtz2015}. Similar sparse structures also emerge in other fields: in natural language processing, where word co-occurrence matrices reveal correlations between contextually related words \cite{Levy2014}; in finance, where asset correlations tend to cluster within sectors \cite{Bouchaud2009}; and in social networks, where relationships between users are captured by sparse covariance matrices \cite{Newman2003}. Additionally, working with large, dense covariance matrices is computationally demanding, often requiring regularisation techniques that induce sparsity and improve efficiency \cite{Fanreview}.

One of the most important observables in the case of random covariance matrix is the top eigenvalue and its associated eigenvector. For instance, in Principal Component Analysis the top eigenvalue and eigenvector capture the most significant variability in data, enabling dimensionality reduction and assisting in signal detection \cite{Vivo2012,johnstone2001,mardia1979,Monasson2015,bai2010,Stocia1996,Bianchi2011,Nadakuditi2008}. 

In this paper, we build on the works \cite{Nagao2007,Susca2019,Susca2020,Susca2021} to formulate a replica approach that is well suited to the average largest eigenvalue and the density of its associated eigenvector's components \color{black} of diluted Wishart matrices. We allow for a large class of weights on non-zero entries that lead to an \emph{isolated} top eigenvalue (see below for more details). 

The outline of the paper is as follows. In section \ref{Literature review} we review the relevant literature; in section \ref{formulation}, we formulate the problem, introduce notations, and specify our assumptions; in section \ref{Replica analysis}, we use the replica formalism to compute the largest eigenvalue of sparse random matrices of the form $\bm{J}=\bm{X}^T\bm{X}$; in Section \ref{sec:replica_eigenvector}, we build on the results of the previous section to compute the density of the corresponding top eigenvector components; \color{black} in Section \ref{PopDyn}, we discuss the population dynamics algorithm used to solve the system of self-consistent equations; in Section \ref{Large q} we show how taking the dense limit recovers the known noncentral Wishart-ensemble results;  \color{black} finally, in Section \ref{Conclusions}, we summarise our results and conclusions; \ref{Upper Bound} provides an upper bound for the largest eigenvalue, which may serve as a suitable starting point for the population dynamics algorithm, while \ref{Average} is devoted to the calculation of a technical average.

\section{Literature review}
\label{Literature review}
\color{black}
Since Wishart's pioneering work \cite{Wishart}, random matrix theory has played a fundamental role in multivariate statistics \cite{Guptabook}. Results derived from random matrix models serve as crucial benchmarks for comparison with empirical data. A central focus of this field is the study of eigenvalue and eigenvector \color{black} statistics, which provide insight into correlations and principal components in complex data.

A common null model for the covariance matrix $\bm{J}=\bm{X}^T\bm{X}$ assumes independent Gaussian random variables, adjusted to have zero mean, as entries of $\bm{X}$. This model yields an analytically known joint distribution of eigenvalues, completely decoupled from the distribution of eigenvectors, \color{black} enabling the application of the Coulomb gas technique in the large $N, M$ limit with their ratio fixed \cite{Dyson1,Dyson2,Dyson3}. This approach has led to extensive results on the eigenvalue statistics of dense covariance matrices \cite{Vivo2007,Isaac2010,Vivo2012,Majumdar2009,Zavatone2023}, including a detailed characterisation of both typical and atypical eigenvalue fluctuations \cite{Vivo2012}. The eigenvectors of this rotationally invariant model are Haar-distributed over the sphere \cite{Forrester2010}, and their associated components follow the Porter-Thomas distribution \cite{Livan2018}. 

\color{black} While the eigenvalue and eigenvector \color{black} statistics of dense covariance matrices are well understood, the situation is markedly different for sparse (``diluted'') covariance matrices, where many entries are zero. Analytical results in this case are primarily limited to the average spectral density \cite{Perez2008,Nagao2007} and the number of eigenvalues in a given interval \cite{Castillo2018}. A key challenge is the absence of an analytical expression for the joint eigenvalue distribution, as the loss of rotational invariance precludes the use of the Coulomb gas approach and other techniques based on orthogonal polynomials \cite{Livan2018} or Fredholm determinants and Painlevé transcendents \cite{forresterpainleve}. While novel methods have expanded our understanding of sparse random matrices \cite{Perez2008,Kuhn2008,Perez2009,Perez2010,MetzA2010,MetzA2011,MetzA2012,MetzA2013,MetzA2014,MetzA2016}, the analytical framework remains less developed compared to the ``classical'' dense case.

A particularly important aspect of covariance matrix spectra is the behaviour of the largest eigenvalue and its associated eigenvector\color{black}, which serves as a key indicator of system-wide correlations. In the \emph{dense} regime, significant progress has been made in characterising the largest eigenvalue distribution. Several works \cite{Peche2007, Karoui2005, Bao2013, Natesh2014, Ding2018} have established that, under fairly general conditions, the largest eigenvalue follows the Tracy-Widom law, demonstrating a form of strong universality. The statistics of eigenvectors of non-centred and doubly correlated Gaussian random matrices has been tackled in \cite{barucca} using a supersymmetric technique.

In contrast, in the \emph{sparse} regime, the largest eigenvalue may exhibit a fundamentally different behaviour, and elementary results remain relatively scarce. While some progress has been made—such as a local Tracy-Widom law for sparse covariance matrices with zero-mean entries \cite{Hwang2019} and studies on heavy-tailed distributions revealing deviations from classical universality \cite{Auffinger2015}—many open questions remain. In particular, for matrices of the form $\bm{X}^T\bm{\Sigma X}$, where $\bm{\Sigma}$ introduces non-uniform sample couplings, the largest eigenvalue can separate from the bulk spectrum. In the dense regime, this phenomenon corresponds to the well-known \emph{BBP transition} \cite{Baik2005}, in which the structure of $\bm{\Sigma}$ drives the detachment. Interestingly, even in the null case, where $\bm{\Sigma}=\mathds{1}$, a spectral gap can emerge if the entries of $\bm{X}$ have a nonzero mean \cite{Bassler2008,Georges2011}. In this scenario, the nonzero-mean entries can be interpreted as a deterministic signal, to which Gaussian noise is added in the form of a zero-mean random covariance matrix. Under this interpretation, the detachment of the largest eigenvalue marks the point at which the noise level becomes low enough to allow inference of the principal component of the original data. Correspondingly, the detachment is accompanied by a similar transition in its associated eigenvector, which becomes localised on a cone whose axis aligns with the principal component of the deterministic signal \cite{Georges2011}. \color{black} In the sparse regime, a qualitatively similar behaviour is observed, but a precise quantitative characterisation of this transition remains an open problem. 

To study spectral properties of large random matrices, various analytical techniques have been developed. Originally introduced in the context of spin glasses \cite{Zamponi2010,BookParisi}, the replica method was first applied to random matrices by Edwards and Jones \cite{Edwards1976} to compute the spectral density of dense matrices. This approach, which relies on the joint distribution of matrix entries rather than eigenvalues, was later extended by Bray and Rodgers \cite{Rodgers1990} to derive an expression for the spectral density of sparse Erdős-Rényi adjacency matrices. However, solving the resulting integral equations remains challenging, with numerical progress made only recently \cite{pipattana2024}.

Alternative functional methods, such as the single defect approximation (SDA) and effective medium approximation (EMA) \cite{Biroli1999, Semerjian2002}, have been developed to tackle these problems. In the context of sparse covariance matrices, these approaches were used in \cite{Nagao2007} to compute the spectral density. Another promising line of research builds on the replica-symmetric framework of Bray and Rodgers, representing order parameters as continuous superpositions of Gaussians with fluctuating variances \cite{Kuhn2008, Bianconi2008}. This method was recently applied in \cite{Susca2019,Susca2020} to study the typical largest eigenvalue of sparse weighted graphs, leading to nonlinear integral equations that can be efficiently solved using a population dynamics algorithm.

These techniques, originally developed in \cite{Kuhn2008,Kuhn2007,Dean2002}, have since found widespread use in random matrix theory \cite{Metz2015,Metz2016B,Metz2017,Lopez2020,sollich2025}, providing a powerful framework for analysing spectral properties beyond the classical setting.

\section{Formulation of the problem}
\label{formulation}
Consider the sparse $N\times M$ matrix $\bm{X}$, whose entries, $X_{ij}$, are random variables, defined as 

\begin{equation}
    X_{ij}=c_{ij}K_{ij}\ .
    \label{eq:X Def}
\end{equation}

\noindent Here, $c_{ij}\in\{0,1\}$ regulates the density of non-zero elements of $\bm{X}$ and $K_{ij}$ represents the non-zero elements' weights, randomly drawn from the pdf $p(K)$. The central object of this study is the $M \times M$ symmetric matrix

\begin{equation}
    \bm J=\bm X^T\bm X\ .
    \label{eq:J Def}
\end{equation}

\noindent We work in the regime $N\rightarrow\infty$ and $M\rightarrow\infty$, but with the ratio
 
 \begin{equation}
     \alpha=\sqrt{\frac{N}{M}}
     \label{eq:alpha Def}
 \end{equation}
 
 \noindent kept finite. According to the spectral theorem, the symmetric matrix $\bm J$ can be diagonalised via an orthonormal basis of eigenvectors, $\{\bm v_m\}_{m=1}^M\in \mathbb{R}^M$, whose corresponding real eigenvalues are denoted by $\{\lambda_m\}_{m=1}^M$. Assuming that the real eigenvalues are not degenerate, we can sort them as $\lambda_1>\lambda_2>...>\lambda_M$. The main goal of this work is to evaluate
 
 \begin{itemize}
     \item The typical value of the largest of them, denoted by $\Big\langle\lambda_1\Big\rangle$ -- assumed to be $\sim\mathcal{O}(1)$ in the limit.
     \item The density of its corresponding eigenvector's components, $T(u)=\Big\langle \frac{1}{M}\sum_{i=1}^M \delta\left(u-v_1^{(i)}\right)\Big\rangle$,
 \end{itemize}
 
 \noindent where $\Big\langle\cdot\Big\rangle$ stands for averaging over different realisations of $\bm X$.

By analogy to the standard prototype of sparse random system (the Erd\H{o}s-R\'enyi graph), the model we study is defined by the following probability to draw (independently) the matrix entries $X_{ij}$

\begin{equation}
    P\left( X_{ij} \right) = \left[\frac{q}{\sqrt{NM}}\delta_{c_{ij},1}+\left(1-\frac{q}{\sqrt{NM}}\right)\delta_{c_{ij},0}\right]p\left(K_{ij}\right)\ .
    \label{eq:X distribution}
\end{equation}

\noindent Indeed, in graph-theoretical terms \cite{Bollobas2001}, the random matrix $\bm{X}$ can be interpreted as the weighted adjacency matrix of a Poissonian bipartite random graph with two distinct node types \cite{Perez2008}: $i$-nodes, corresponding to the rows of $\bm{X}$, and $j$-nodes, corresponding to its columns. The matrix $\bm X$ is sparse in the sense that the average number $q/\alpha$ of its non-zero elements per row does not scale with either $N$ or $M$. 

The model defined in \eqref{eq:X distribution} suffers from two potential drawbacks, though: (i) without further restrictions on the \emph{maximal} number of nonzero elements allowed in each row and column, the largest eigenvalue may (slowly) grow with $N,M$, in contrast with our assumption that 
$\Big\langle\lambda_1\Big\rangle\sim\mathcal{O}(1)$; and (ii) a general and unrestricted weight distribution $p(K)$ may lead to a largest eigenvalue that is not detached from the continuous bulk of the spectrum. 

The concern (i) follows from the observation that -- in the similar case of (square) adjacency matrices of sparse random graphs -- the largest eigenvalue (without further restrictions) indeed grows (slowly) with $N$, as proven in \cite{Krivelevich2001}. Although we are not aware of a similar theorem in the context of diluted correlation matrices, it is a plausible assumption that a similar mechanism may be at work here. To ensure that the largest eigenvalue remains $\mathcal{O}(1)$, we therefore impose the constraints that there be at most $C$ nonzero elements per column and $R$ per row. In \ref{Upper Bound} we show that such a constraint indeed results in an $\mathcal{O}(1)$ upper bound for $\langle \lambda_1 \rangle$. 

The concern (ii) can be allayed more easily by assuming that the weight pdf $p(K)$ is such that the largest eigenvalue is isolated, i.e. there is a macroscopic gap between it and the sea of smaller eigenvalues. This detachment also occurs in the dense regime, where precise relationships between $p(K)$ and the resulting spectral gap can be established \cite{Bassler2008} -- notably, a necessary (though not sufficient) condition is that $p(K)$ has a non-zero mean. The sparse regime exhibits a similar qualitative behaviour, although a complete analytical characterisation of the transition remains an open problem. Throughout the rest of the paper, when referring to $p(K)$ as a `non-zero mean distribution', we specifically mean it in the sense of it generating a spectral gap.

\color{black}

While the restriction on the maximal number of nonzero elements in $\bm{X}$ imposes nontrivial couplings between the $c_{ij}$'s, which would in principle require a re-working of the form \eqref{eq:X distribution} of the pdf of entries and introduce an additional analytical burden, we benefit here from a key observation made in \cite{Kuhn2008,Kuhn2015,Susca2019}: a convenient shortcut for the calculation consists in (i) initially replacing the ``microcanonical'' version of the model with the simpler ``canonical'' one, in which the $c_{ij}$'s are \emph{independent} Bernoulli random variables with success probability $q/\sqrt{NM}$ (as given in \eqref{eq:X distribution}), and (ii) manually adjusting the Poissonian distribution of ``degrees'' of the connectivity matrix -- which naturally emerges in the replica calculation -- to allow for a finite maximum number of nonzero entries per row and column [See, e.g. Eqs. \eqref{eq:rhoSaddle_Final} to \eqref{eq:IntegralCondition_Final} and discussion after Eq. \eqref{eq: Lambda Lambda}]. For all the technical details that motivate this shortcut, we refer to Appendix B in \cite{Susca2019}. 

In the next section, we provide a detailed analysis of the replica calculation for the typical largest eigenvalue of diluted Wishart matrices.

\section{Replica analysis of the typical largest eigenvalue} 
\label{Replica analysis}

We begin our analysis by noting that the problem of evaluating $\bm{J}$'s largest eigenvalue can be formulated in terms of the Courant-Fisher maximisation

\begin{equation}
    \lambda_1 = \frac{1}{M}\max_{\bm v\in \mathbb{R}^M,\ |\bm v|^2=M} \left\langle\bm v,\bm J \bm v\right\rangle\ ,
    \label{eq:Lambda1 Opt}
\end{equation}

\noindent where $\Big\langle\cdot,\cdot\Big\rangle$ stands for the standard dot product among vectors in $\mathbb{R}^M$. We now introduce an auxiliary canonical partition function at inverse temperature $\beta$

\begin{equation}
     Z=\int \mathrm{d}\bm v~ \exp\left(\frac{\beta}{2}\Big\langle\bm v,\bm J\bm v\Big\rangle\right)\delta\left(|\bm v|^2-M\right)\ .\label{canonicalpart}
\end{equation}
In the zero-temperature limit $\beta\to\infty$, applying the Laplace method to the integral in \eqref{canonicalpart} we obtain using \eqref{eq:Lambda1 Opt} that
\begin{equation}
    Z\approx \exp\left(\frac{\beta}{2}\max_{\bm v\in \mathbb{R}^M,\ |\bm v|^2=M} \left\langle\bm v,\bm J \bm v\right\rangle\right)=\exp\left(\frac{\beta}{2}M\lambda_1\right)\ .
\end{equation}

\noindent Therefore

\begin{equation}
    \Big\langle\lambda_1\Big\rangle = \lim_{\beta\rightarrow\infty} \frac{2}{\beta M} \Big\langle \ln Z\Big\rangle\ . 
    \label{eq:Lambda1 StatMech}
\end{equation}

\noindent To tackle the average on the r.h.s of Eq. \eqref{eq:Lambda1 StatMech} we invoke the replica trick \cite{BookParisi}

\begin{equation}
    \Big\langle\lambda_1\Big\rangle = \lim_{\beta\rightarrow\infty} \frac{2}{\beta M}  \lim_{n\rightarrow 0} \frac{1}{n}\ln\Big\langle Z^n \Big\rangle\ ,
    \label{eq:Lambda_Replica}
\end{equation}

\noindent where $n$ is initially treated as an integer, and then analytically continued to real values around $n=0$. The next section is devoted to computing the average of the replicated partition function $\Big\langle Z^n\Big\rangle$.\\

The evaluation of \eqref{eq:Lambda_Replica} proceeds through five principal steps: (1) We average the replicated partition function over the disorder (randomness encoded in the matrix entries $X_{ij}$); (2) We reformulate the replicated partition function in terms of a functional integral, which lends itself to a suitable form for a saddle point analysis; (3) To study the saddle point structure of the replicated partition function, we then invoke a replica symmetric ansatz that recasts the order parameters as a superposition of an uncountably infinite set of Gaussians; (4) We use the replica symmetric ansatz to derive the saddle point equations, and finally, (5) We evaluate the replicated partition function at the saddle point in the limit \(\beta\rightarrow\infty\). 

We will guide the reader through the various steps below. 

\subsection{Averaging the replicated partition function over the disorder}
\label{Exponentiating}
\color{black}

The first step in our analysis is to write the replicated partition function,

\begin{equation}
\Big\langle Z^n\Big\rangle = \int \left(\prod_{a=1}^n  \mathrm{d}\bm{v}_a\right) \Big\langle\exp\left[ \frac{\beta}{2} \sum_{a=1}^n \sum_{i,k=1}^M v_{ia} J_{ik} v_{ka} \right]\Big\rangle \prod_{a=1}^n  \delta\left(|\bm{v}_a|^2-M\right)\ ,
\label{eq:Zn}
\end{equation}

\noindent as an integral of an exponential function. Expressing $J_{ik}=\sum_{j=1}^N X_{ji}X_{jk}$, we note that

\begin{align}
\nonumber    &\Big\langle\exp \left[ \frac{\beta}{2} \sum_{a=1}^n \sum_{i,k=1}^M v_{ia} J_{ik} v_{ka} \right]\Big\rangle \nonumber = \Big\langle\exp\left[ \frac{\beta}{2} \sum_{a=1}^n \sum_{i,k=1}^M \sum_{j=1}^N X_{ji}v_{ia} X_{jk}v_{ka} \right]\Big\rangle \nonumber \\
    &=\Big\langle\prod_{a=1}^n \prod_{j=1}^N \exp\left[ \frac{\beta}{2}\left( \sum_{i=1}^M v_{ia}X_{ji}\right)^2\right] \Big\rangle \nonumber =\Big\langle\prod_{a=1}^n \prod_{j=1}^N  \sqrt{\frac{\beta}{2\pi}} \int \mathrm{d}u \exp\left(-\frac{\beta}{2}u^2+\beta u\sum_{i=1}^M v_{ia}X_{ji}\right) \Big\rangle \nonumber \\
    &=\left(\frac{\beta}{2\pi}\right)^{\frac{Nn}{2}} \int \left(\prod_{a=1}^n \prod_{j=1}^N \mathrm{d}u_{ja} \mathrm{e}^{-\frac{\beta}{2}u_{ja}^2} \right) \Big\langle \prod_{i=1}^M \prod_{j=1}^N \exp\left(\beta X_{ji}\sum_{a=1}^n v_{ia}u_{ja}\right)\Big\rangle\ ,
    \label{eq:ExpAv}
\end{align}
     
\noindent where we used the Hubbard-Stratonovich transformation,

\begin{equation}
    \int_{-\infty}^\infty \mathrm{d}x~\mathrm{e}^{-ax^2+bx}=\sqrt{\frac{\pi}{a}}\mathrm{e}^{\frac{b^2}{4a}}\ .
\end{equation}

\noindent In \ref{Average} we show that using the sparsity condition, the average in Eq. \eqref{eq:ExpAv} can be performed for large $N,M$ as

\begin{align}
         \Big\langle \prod_{i=1}^M \prod_{j=1}^N &\exp\left(\beta X_{ji}\sum_{a=1}^n v_{ia}u_{ja}\right)\Big\rangle\nonumber \\
         &\simeq \exp\left\{ \frac{q}{\sqrt{NM}}\sum_{i=1}^M\sum_{j=1}^N \left[\Big\langle\exp\left(\beta K\sum_{a=1}^n v_{ia}u_{ja}\right)\Big\rangle_K -1\right] \right\}\ ,
        \label{eq:J Average}
\end{align}

\noindent where the average $\langle\cdot\rangle_K$ is over a single realisation of the random variable $K$ drawn from $p(K)$, the weight distribution. Furthermore, we use the Fourier representation of the delta function,

\begin{equation}
    \prod_{a=1}^n \delta\left(|\bm{v}_a|^2 - M\right) = \int_{-\infty}^\infty \left(\prod_{a=1}^n \frac{\beta}{2}\frac{\mathrm{d}\lambda_a}{2\pi}\right) \prod_{a=1}^n \exp\left[ -\mathrm{i}\frac{\beta}{2}\lambda_a\left(\sum_{i=1}^M v_{ia}^2 -M\right) \right],
    \label{eq:Delta integral repres}
\end{equation}

\noindent such that Eq. \eqref{eq:Zn} takes the form (ignoring pre-factors whose logarithm vanishes in the limit) 
\color{black}

\begin{align}
    &\Big\langle Z^n\Big\rangle \propto \int \left( \prod_{a=1}^n \mathrm{d}\bm{v}_a \mathrm{d}\bm{u}_a \mathrm{d}\lambda_a\right) \exp\left(-\frac{\beta}{2}\sum_{a=1}^n \sum_{j=1}^N u_{ja}^2\right) \exp\left(\mathrm{i}M \frac{\beta}{2}\sum_{a=1}^n \lambda_a\right) \times \nonumber \\
    & \times \exp\left(-\mathrm{i} \frac{\beta}{2}\sum_{a=1}^n\sum_{i=1}^M \lambda_a v_{ia}^2\right) \exp\left\{ \frac{q}{\sqrt{NM}}\sum_{i=1}^M\sum_{j=1}^N \left[\Big\langle\exp\left(\beta K\sum_{a=1}^n v_{ia}u_{ja}\right)\Big\rangle_K -1\right] \right\}\ .
    \label{eq:ZnExplic}
\end{align}
Note that in \eqref{eq:ZnExplic}, $\{\bm u_a\}_{a=1}^n \in \mathbb{R}^N$ and $\{\bm v_a\}_{a=1}^n \in \mathbb{R}^M$.

\subsection{Functional integral representation}
\label{Functional Integral Representation}
\color{black}
\noindent Next, we aim at expressing the replicated partition function through a functional integral over the following order parameters

\begin{gather}
    \phi(\Vec{v})=\frac{1}{M} \sum_{i=1}^M \prod_{a=1}^n \delta\left( v_a - v_{ia} \right) \label{eq:RepDensPhi} \\
    \psi(\Vec{u})=\frac{1}{N} \sum_{j=1}^N \prod_{a=1}^n \delta\left( u_a - u_{ja} \right)\ ,
    \label{eq:RepDensPsi}
\end{gather}

\noindent where $\vec{v},\vec{u}\in \mathbb{R}^n$ are $n$-dimensional vectors in replica space. The order parameters were chosen as such since this approach will eventually lead to a symmetric representation of the replicated partition function under the duality transformation $\alpha\rightarrow1/\alpha$. This symmetry reflects the simple fact that the matrix $\bm J=\bm X^T\bm X$ shares its largest eigenvalue with its `dual' $N\times N$ counterpart $\tilde{\bm J}=\bm X\bm X^T$. This approach serves as a starting point for a functional scheme introduced in \cite{Nagao2007}
for the analysis of the spectral density of $\bm{J}$. 

To enforce the definitions given in \cref{eq:RepDensPhi,eq:RepDensPsi} upon the replicated partition function, we multiply Eq. \eqref{eq:ZnExplic} by the functional-integral representations of the identity

\begin{gather}
    1 = \int M \mathcal{D}\phi \mathcal{D}\hat{\phi} \exp \left\{ -\mathrm{i}\int \mathrm{d}\vec{v}\hat{\phi}\left(\vec{v}\right) \left[ M\phi\left(\vec{v}\right) - \sum_{i=1}^M \prod_{a=1}^n \delta\left( v_a - v_{ia} \right) \right] \right\}  \\ 
    1 = \int N \mathcal{D}\psi \mathcal{D}\hat{\psi} \exp \left\{ -\mathrm{i}\int \mathrm{d}\vec{u}\hat{\psi}\left(\vec{u}\right) \left[ N\psi\left(\vec{u}\right) - \sum_{j=1}^N \prod_{a=1}^n \delta\left( u_a - u_{ja} \right) \right] \right\}\ ,
\end{gather}

\noindent where $\mathrm{d}\vec{v}= \prod_{a=1}^n \mathrm{d}v_a$, and similarly $\mathrm{d}\vec{u}= \prod_{a=1}^n \mathrm{d}u_a$. \color{black} This allows us to rewrite Eq. \eqref{eq:ZnExplic} as

\begin{align}
     \Big\langle Z^n\Big\rangle &\propto \int \mathcal{D}\phi \mathcal{D}\hat{\phi} \mathcal{D}\psi \mathcal{D}\hat{\psi}\mathrm{d}\vec{\lambda} \exp\left[ -\mathrm{i}M \int \mathrm{d}\vec{v} \hat{\phi}\left(\vec{v}\right) \phi\left(\vec{v}\right) -\mathrm{i}N \int \mathrm{d}\vec{u} \hat{\psi}\left(\vec{u}\right) \psi\left(\vec{u}\right) \right] \nonumber \\
     &\times \exp \left[ q\sqrt{NM}\int \mathrm{d}\vec{v}\mathrm{d}\vec{u}\phi\left(\vec{v}\right)\psi\left(\vec{u}\right)\left(\Big\langle \mathrm{e}^{\beta K \vec{v}\cdot\vec{u}}\Big\rangle_K-1\right) + \mathrm{i}M \frac{\beta}{2}\sum_{a=1}^n\lambda_a \right] \nonumber \\
     &\times    \int \left( \prod_{a=1}^n \mathrm{d}\bm{v}_a \right) \exp\left[-\frac{\beta}{2} \sum_{a=1}^n\sum_{i=1}^M \lambda_a v_{ia}^2 + \mathrm{i}\sum_{i=1}^M\int \mathrm{d}\vec{v} \hat{\phi}\left(\vec{v}\right)\prod_{a=1}^n \delta\left(v_a-v_{ia}\right) \right] \nonumber \\
     &\times  \int \left( \prod_{a=1}^n \mathrm{d}\bm{u}_a \right) \exp\left[-\frac{\beta}{2} \sum_{a=1}^n \sum_{j=1}^N u_{ja}^2 + \mathrm{i}\sum_{j=1}^N\int \mathrm{d}\vec{u} \hat{\psi}\left(\vec{u}\right)\prod_{a=1}^n \delta\left(u_a-u_{ja}\right) \right]\ .
     \label{eq:Zn Med}
\end{align}

\noindent Note that the two multiple integrals appearing in the last two lines of Eq. \eqref{eq:Zn Med} can be factorised into $M$ and $N$ identical $n$-fold integrals respectively, 

\begin{align}
    I_M &=\int \left( \prod_{a=1}^n \mathrm{d}\bm{v}_a \right) \exp\left[-\frac{\beta}{2} \sum_{a=1}^n\sum_{i=1}^M \lambda_a v_{ia}^2 + \mathrm{i}\sum_{i=1}^M\int \mathrm{d}\vec{v} \hat{\phi}\left(\vec{v}\right)\prod_{a=1}^n \delta\left(v_a-v_{ia}\right) \right] \nonumber \\ &= \left\{ \int \mathrm{d}\vec{v} \exp\left[ -\mathrm{i} \frac{\beta}{2}\sum_{a=1}^n \lambda_a v_{a}^2 + \mathrm{i}\hat{\phi}\left(\vec{v}\right) \right] \right\}^M ,
\end{align}
\begin{align}
        I_N&=\int \left( \prod_{a=1}^n \mathrm{d}\bm{u}_a \right) \exp\left[-\frac{\beta}{2} \sum_{a=1}^n\sum_{j=1}^N u_{ja}^2 + \mathrm{i}\sum_{j=1}^N\int \mathrm{d}\vec{u} \hat{\psi}\left(\vec{u}\right)\prod_{a=1}^n \delta\left(u_a-u_{ja}\right) \right] \nonumber \\ &= \left\{ \int \mathrm{d}\vec{u} \exp\left[ -\mathrm{i} \frac{\beta}{2}\sum_{a=1}^n  u_{a}^2 + \mathrm{i}\hat{\psi}\left(\vec{u}\right) \right] \right\}^N\ ,    
\end{align}

\noindent such that \eqref{eq:Zn Med} can be written as

\begin{equation}
   \Big\langle Z^n \Big\rangle\propto\int \mathcal{D}\phi \mathcal{D}\hat{\phi} \mathcal{D}\psi \mathcal{D}\hat{\psi} \mathrm{d}\vec{\lambda} ~\mathrm{e}^{\sqrt{NM}S\left[ \phi,\hat{\phi},\psi,\hat{\psi};\vec{\lambda} \right]}\ .
\label{eq:ZnPathIntegral}
\end{equation}

\noindent The action $S\left[ \phi,\hat{\phi},\psi,\hat{\psi};\vec{\lambda} \right]$ is defined as

\begin{equation}
     S\left[ \phi,\hat{\phi},\psi,\hat{\psi};\vec{\lambda} \right] = S_1\left[ \phi,\hat{\phi}\right] +  S_2 \left[\hat{\phi};\vec{\lambda}\right] +\Tilde{S}_1 \left[ \psi,\hat{\psi}\right] + \tilde{S}_2 \left[\hat{\psi} \right] + S_3 \left[\vec{\lambda} \right] + S_{int}\left[ \phi,\psi\right]\ ,
     \label{eq:S_imp}
\end{equation}

\noindent where

\begin{gather}
    S_1\left[ \phi,\hat{\phi}\right] = -\frac{\mathrm{i}}{\alpha} \int \mathrm{d}\vec{v} \hat{\phi}\left(\vec{v}\right) \phi\left(\vec{v}\right) \label{eq:S1} \\
    S_2 \left[\hat{\phi};\vec{\lambda}\right] = \frac{1}{\alpha} \Log\int \mathrm{d}\vec{v} \exp\left[ -\mathrm{i} \frac{\beta}{2} \sum_{a=1}^n \lambda_a v_a^2 +\mathrm{i}\hat{\phi}\left(\vec{v}\right)\right] \label{eq:S2} \\
    \tilde{S}_1 \left[ \psi,\hat{\psi}\right] = -\mathrm{i}\alpha  \int \mathrm{d}\vec{u} \hat{\psi}\left(\vec{u}\right) \psi\left(\vec{u}\right) \label{eq:S1_tilde} \\
    \tilde{S}_2 \left[\hat{\psi} \right] = \alpha~\Log\int \mathrm{d}\vec{u} \exp\left[ -\mathrm{i} \frac{\beta}{2} \sum_{a=1}^n u_a^2 +\mathrm{i}\hat{\psi}\left(\vec{u}\right)\right] \label{eq:S2_tilde} \\
    S_3 [\vec{\lambda}]= \mathrm{i}\frac{\beta}{2\alpha}\sum_{a=1}^n \lambda_a \label{eq:S3} \\
    S_{int}\left[ \phi,\psi\right] = q\int \mathrm{d}\vec{v} \mathrm{d}\vec{u} \phi\left(\vec{v}\right) \psi\left(\vec{u}\right)\left(\Big\langle \mathrm{e}^{\beta K \vec{v}\cdot\vec{u}}\Big\rangle_K-1\right) \label{eq:Sint}\ , 
\end{gather}

\noindent and $\Log$ is the branch of the complex logarithm such that $\Log~\mathrm{e}^z = z$. 

{The form \eqref{eq:ZnPathIntegral} is amenable to a saddle-point evaluation for large $N,M$. In order to facilitate the $n\to 0$ limit, we will first adopt a replica symmetric ansatz as detailed in the sub-section below.}

\subsection{Replica-symmetric ansatz} \label{replicasymmetricansatz}

We now employ a replica symmetric ansatz, \color{black} which assumes that the dependence on the vector arguments $\vec v$ and $\vec u$ is only through a permutation-symmetric function of the vector components. An even stronger ``rotationally invariant'' assumption -- namely that such dependence would only be through the modulus $|\vec v|$ and $|\vec u|$ of the vectors involved -- was shown to lead to the correct solution for the \emph{spectra} of sparse random matrices \cite{Edwards1976,Rodgers1990,Kuhn2008,BookParisi}. However, for questions related to the largest eigenvalue/eigenvector, the latter assumption was shown to be too restrictive on the space of function within which to seek for an extremiser of the action \cite{Nagao2007,Susca2019,Susca2020,Susca2021}.

The permutation-symmetric ansatz consists in writing the replicated order parameters as a superposition of uncountably infinite Gaussians with non-zero mean. We will follow this prescription, as originally suggested in \cite{Kuhn2007,Kuhn2008,Dean2002}, while noting that it is not the most general possible as it does not include cross-terms. 

To this end, we introduce the following normalised densities, $\pi(\omega,h)$, $\hat{\pi}(\hat{\omega},\hat{h})$, $\rho(\sigma,\mu)$, $\hat{\rho}(\hat{\sigma}$, $\hat{\mu})$, and their respective measures, $\mathrm{d}\pi = \mathrm{d}\omega~\mathrm{d}h~\pi(\omega,h)$, $\mathrm{d}\hat{\pi}=\mathrm{d}\hat{\omega}~\mathrm{d}\hat{h} ~\hat{\pi}(\hat{\omega},\hat{h})$, $\mathrm{d}\rho = \mathrm{d}\sigma \mathrm{d}\mu\rho(\sigma,\mu)$ and $\mathrm{d}\hat{\rho}=\mathrm{d}\hat{\sigma}\mathrm{d}\hat{\mu} \hat{\rho}(\hat{\sigma},\hat{\mu})$. We then use these densities to represent the replicated order parameters as 

\begin{gather}
    \phi\left(\vec{v}\right) = \int \mathrm{d}\pi \prod_{a=1}^n \frac{1}{\mathcal{Z}_\beta(\omega,h)}\mathrm{e}^{-\frac{\beta}{2}\omega v_a^2 + \beta h v_a} \label{eq:phiGauss} \\ 
    \mathrm{i}\hat{\phi}\left(\vec{v}\right) = \hat{c}\int \mathrm{d}\hat{\pi} \prod_{a=1}^n \mathrm{e}^{\frac{\beta}{2}\hat{\omega} v_a^2 + \beta \hat{h} v_a} \label{eq:phi_hatGauss} \\ 
    \psi\left(\vec{u}\right) = \int \mathrm{d}\rho \prod_{a=1}^n \frac{1}{\mathcal{Z}_\beta(\sigma,\mu)}e^{-\frac{\beta}{2}\sigma u_a^2 + \beta \mu u_a} \label{eq:psiGauss}\\    
    \mathrm{i}\hat{\psi}\left(\vec{u}\right) = \hat{t}\int \mathrm{d}\hat{\rho} \prod_{a=1}^n \mathrm{e}^{\frac{\beta}{2}\hat{\sigma} u_a^2 + \beta \hat{\mu} u_a} \label{eq:psi_hatGauss}\\ 
    \mathrm{i}\lambda_a = \lambda \quad\quad \forall\ 1\leq a\leq n \label{eq:LambdaGauss}\ ,
\end{gather}

\noindent with

\begin{equation}
    \mathcal{Z}_\beta(x,y) = \sqrt{\frac{2\pi}{\beta x}} \mathrm{e}^{\frac{\beta y^2}{2x}}.
    \label{eq:Zbeta}
\end{equation}

\noindent Note that since $\pi,\hat{\pi},\rho$ and $\hat{\rho}$ are normalised densities, this representation preserves the normalisation of $\phi\left(\vec{v}\right)$ and $\psi\left(\vec{u}\right)$. The constants $\hat c$ and $\hat t$ are introduced to account for the fact that the conjugate functions $\mathrm{i}\hat\phi$ and $\mathrm{i}\hat\psi$ do not have the interpretation of a density, therefore they need not be normalised. 

This representation allows us to integrate out the $\vec{v}$'s and $\vec{u}$'s and extract the leading $n\rightarrow 0$ behaviour, which is currently only implicit in \eqref{eq:S_imp} (for full details of how to apply the transformation, see appendix E in \cite{Susca2021}). \color{black}Inserting \cref{eq:phiGauss,eq:phi_hatGauss,eq:psiGauss,eq:psi_hatGauss,eq:LambdaGauss} into \cref{eq:S1,eq:S2,eq:S1_tilde,eq:S2_tilde,eq:S3,eq:Sint} and collecting terms up to $\mathcal{O}(n)$, while introducing Lagrange multipliers that enforce normalisation upon the densities, the action takes the form of

\begin{align}
    S\left[\pi, \hat{\pi}, \rho, \hat{\rho}; \lambda\right]\simeq&-\frac{n\hat{c}}{\alpha}\int \mathrm{d}\pi \ \mathrm{d}\hat{\pi} \Log\left[\frac{\mathcal{Z}_\beta\left(\omega-\hat{\omega}, h+\hat{h}\right)}{\mathcal{Z}_\beta\left(\omega, h\right)}\right] \nonumber \\
    &+ \frac{n}{\alpha}\sum_{s=0}^\infty p_{\hat{c}}\left(s\right)\int \{\mathrm{d}\hat{\pi}\}_s\Log \mathcal{Z}_\beta\left(\lambda-\{\hat{\omega}\}_s,\{\hat{h}\}_s\right) \nonumber
    \\
    & -n\hat{t}\alpha\int \mathrm{d}\rho \mathrm{d}\hat{\rho} \Log\left[\frac{\mathcal{Z}_\beta\left(\sigma-\hat{\sigma}, \mu +\hat{\mu}\right)}{\mathcal{Z}_\beta\left(\sigma,\mu\right)}\right] \nonumber \\
    &+ n\alpha\sum_{s=0}^\infty p_{\hat{t}}\left(s\right)\int\{\mathrm{d}\hat{\rho}\}_s\Log \mathcal{Z}_\beta\left(1-\{\hat{\sigma}\}_s,\{\hat{\mu}\}_s\right) \nonumber
    \\
    & +n\frac{\beta}{2\alpha}\lambda +nq\int \mathrm{d}\pi \mathrm{d}\rho \Big\langle\Log \left[\frac{\mathcal{Z}_\beta\left(\omega -\frac{K^2}{\sigma}, h +\frac{K\mu}{\sigma}\right)}{\mathcal{Z}_\beta\left(\omega, h\right)}\right]\Big\rangle_K \nonumber \\
    &+\gamma\left(\int\mathrm{d}\pi-1\right) +\hat{\gamma}\left(\int\mathrm{d}\hat{\pi}-1\right)
    +\xi\left(\int\mathrm{d}\rho-1\right) +\hat{\xi}\left(\int\mathrm{d}\hat{\rho}-1\right)\ ,
    \label{eq:SGauss}
\end{align}

\color{black}
\noindent where we introduced the shorthands $\{\mathrm{d}\hat{\pi}\}_s = \prod_{\ell=1}^s\mathrm{d}\hat{\pi}_\ell$, $\{\hat{\omega}\}_s = \sum_{\ell=1}^s\hat{\omega}_\ell$, $\{\hat{h}\}_s = \sum_{\ell=1}^s\hat{h}_\ell$, and similarly with $\hat{\rho}$, $\hat{\sigma}$ and $\hat{\mu}$. Moreover, we denoted by $p_m(s) = \mathrm{e}^{-m}m^s/s!$ the Poisson distribution with mean $m$. Note that for the $\vec{u}$ and $\vec{v}$ integrals to converge, one has to formally require the following inequalities, $\omega>\hat{\omega}$, $\omega>0$, $\lambda > \{\hat{\omega}\}_s$ and similarly, $\sigma>\hat{\sigma}$, $\sigma>0$, $1 > \{\hat{\sigma}\}_s$. Furthermore, if we denote the lower (upper) bound of the support of $p(K)$ by $\zeta^-$ ($\zeta^+)$, another requirement is $\omega\sigma>\left[\max\left(|\zeta^-|,|\zeta^+|\right)\right]^2$. In practice, to satisfy these constraints, one has to dynamically enforce them while running the population dynamics algorithm (see Section \ref{PopDyn}). 

\subsection{Saddle point analysis}
\label{Saddle Point analysis}

We now proceed with our fourth step, which involves studying the saddle point structure of the normalised densities introduced in \ref{replicasymmetricansatz} under the replica-symmetric framework.  
\color{black}
In the limit of $N,M\rightarrow \infty$, Eq. \eqref{eq:ZnPathIntegral} is evaluated using a saddle-point method to give

\begin{equation}
   \Big\langle Z^n \Big\rangle \approx \mathrm{e}^{\sqrt{NM}S\left[ \pi^\star,\hat{\pi}^\star,\rho^\star,\hat{\rho}^\star;\lambda^\star \right]}\ ,
   \label{eq:ZnSaddleImplicit}
\end{equation}

\noindent where $\pi^\star,\hat{\pi}^\star,\rho^\star,\hat{\rho}^\star$ are the saddle point forms of the densities, obtained from the stationary conditions $\delta S/\delta\pi\vert_{\pi^\star,\hat{\pi}^\star,\rho^\star,\hat{\rho}^\star;\lambda^\star}=0$ and similar, and `$\approx$' denotes equivalence on a logarithmic scale. To facilitate the notation, from now on we discard the $\star$'s when addressing the saddle point forms of the densities. Consequently, the first stationary condition, $\delta S/\delta\pi=0$, entails

\begin{equation}
    \frac{\hat{c}}{\alpha q} \int \mathrm{d}\hat{\pi} \Log\left[\frac{\mathcal{Z}_\beta\left(\omega-\hat{\omega}, h+\hat{h}\right)}{\mathcal{Z}_\beta\left(\omega, h\right)}\right] = \int \mathrm{d}\rho \Big\langle\Log\left[\frac{\mathcal{Z}_\beta\left(\omega -\frac{K^2}{\sigma}, h +\frac{K\mu}{\sigma}\right)}{\mathcal{Z}_\beta\left(\omega, h\right)}\right]\Big\rangle_K +\frac{\gamma}{q}\ ,
    \label{eq:Saddle inter}
\end{equation}

\noindent where $\gamma$ is the Lagrange multiplier enforcing the normalisation of $\pi$. To match the two sides of Eq. (\ref{eq:Saddle inter}) for all values of the non-integrated variables, $\omega$ and $h$ \cite{Susca2021}, while preserving normalisation of $\hat{\pi}$, we set 
\begin{gather}
    \hat{\pi}(\hat{\omega},\hat{h}) = \int \mathrm{d}\rho \Big\langle \delta\left( \hat{\omega}-\frac{K^2}{\sigma}\right)\delta\left( \hat{h}- \frac{K\mu}{\sigma}\right)\Big\rangle_K
    \label{eq:pi_hatSaddle} \\
    \hat{c}=\alpha q \label{eq:c_hatSaddle} \\
    \gamma=0\ .
\end{gather}

\noindent To obtain the next stationary condition, $\delta S/\delta\rho=0$, we first note that the interaction term in \eqref{eq:SGauss} was evaluated by integrating out first the $u$'s and then the $v$'s. However, one could have equally well swapped the order of integrations, which results in an equivalent form of $S_{int}$ given by

\begin{equation}
    S_{int}\left[\pi,\rho \right] = n q \int \mathrm{d}\pi \mathrm{d}\rho \Big\langle \Log \left[ \frac{\mathcal{Z}_\beta \left( \sigma-\frac{K^2}{\omega},\mu + \frac{Kh}{\omega} \right)}{\mathcal{Z}_\beta(\sigma,\mu)} \right] \Big\rangle_K\ .
    \label{eq:SintDual}
\end{equation}

\noindent Keeping that in mind, the stationary condition $\delta S/\delta\rho=0$ can be written as 

\begin{equation}
    \frac{\alpha \hat{t}}{q} \int \mathrm{d}\hat{\rho} \Log\left[ \frac{\mathcal{Z}_\beta(\sigma-\hat{\sigma},\mu+\hat{\mu})}{\mathcal{Z}_\beta(\sigma,\mu)} \right] = \int \mathrm{d}\pi \Big\langle \Log \left[ \frac{\mathcal{Z}_\beta \left( \sigma-\frac{K^2}{\omega},\mu + \frac{Kh}{\omega} \right)}{\mathcal{Z}_\beta(\sigma,\mu)} \right] \Big\rangle_K +\frac{\xi}{q}\ ,
\end{equation}

\noindent where $\xi$ is the Lagrange multiplier enforcing normalisation of $\rho$. Using the same argument that led us to Eq. \eqref{eq:pi_hatSaddle}, we find that

\begin{gather}
    \hat{\rho}\left(\hat{\sigma},\hat{\mu}\right) = \int \mathrm{d}\pi \Big\langle \delta\left( \hat{\sigma}-\frac{K^2}{\omega}\right) \delta\left( \hat{\mu}- \frac{Kh}{\omega}\right)\Big\rangle_K \label{eq:rho_hatSaddle}\\
    \hat{t}=\alpha^{-1}q \label{eq:t_hatSaddle}\\
    \xi = 0\ .
\end{gather}

\noindent The next stationary condition, $\delta S/\delta\hat{\pi}=0$, is given by 

\begin{align}
    \int \mathrm{d}\pi &\Log\left[\frac{\mathcal{Z}_\beta\left(\omega-\hat{\omega}, h+\hat{h}\right)}{\mathcal{Z}_\beta\left(\omega, h\right)}\right] \nonumber \\ &= \sum_{s=0}^\infty \frac{s p_{\hat{c}}(s)}{\hat{c}} \int \left\{ \mathrm{d}\hat{\pi} \right \}_{s-1} \Log \mathcal{Z}_\beta\left(\lambda-\{\hat{\omega}\}_{s-1} -\hat{\omega},\{\hat{h}\}_{s-1} +\hat{h}\right) +\frac{\hat{\gamma}}{\hat{c}}\ ,
\end{align}

\noindent where $\hat{\gamma}$ is the Lagrange multiplier enforcing normalisation of $\hat{\pi}$. Using $\hat{c}=\alpha q$ [Eq. \eqref{eq:c_hatSaddle}] we thus find that 

\begin{gather}
    \pi\left(\omega, h\right)=\sum_{s=1}^\infty\frac{sp_{\alpha q}(s)}{\alpha q}\int \{\mathrm{d}\hat{\pi}\}_{s-1}\delta\left(\omega-\left(\lambda-\{\hat{\omega}\}_{s-1}\right)\right)\delta\left(h-\{\hat{h}\}_{s-1}\right) \label{eq:piSaddle} \\
    \hat{\gamma}=-\alpha q \int \mathrm{d}\pi \Log \mathcal{Z}_\beta \left(\omega,h\right)\ .
\end{gather}

\noindent The next stationary condition, $\delta S/\delta\hat{\rho}=0$, reads

\begin{align}
    \int \mathrm{d}\rho &\Log\left[\frac{\mathcal{Z}_\beta\left(\sigma-\hat{\sigma}, \mu +\hat{\mu}\right)}{\mathcal{Z}_\beta\left(\sigma,\mu\right)}\right] \nonumber \\
    &= \sum_{s=0}^\infty \frac{s p_{\hat{t}}(s)}{\hat{t}} \int \left\{ \mathrm{d}\hat{\rho} \right \}_{s-1} \Log \mathcal{Z}_\beta\left(1-\{\hat{\sigma}\}_{s-1} -\hat{\sigma},\{\hat{\mu}\}_{s-1} +\hat{\mu}\right) +\frac{\hat{\xi}}{\hat{t}}\ ,
\end{align}

\noindent where $\hat{\xi}$ is the Lagrange multiplier enforcing normalisation of $\hat{\rho}$. Using $\hat{t}=\alpha^{-1}q$ [Eq. \eqref{eq:t_hatSaddle}], the saddle point form of $\rho$ can be expressed as

\begin{gather}
    \rho\left( \sigma,\mu \right) = \sum_{s=1}^\infty \frac{sp_{\alpha^{-1}q}}{\alpha^{-1}q} \int \left\{ \mathrm{d}\hat{\rho} \right\}_{s-1} \delta\left(\sigma-\left(1-\{\hat{\sigma}\}_{s-1}\right)\right)\delta\left(\mu-\{\hat{\mu}\}_{s-1}\right) \label{eq:rhoSaddle} \\
     \hat{\xi}=-\alpha^{-1} q \int \mathrm{d}\rho \Log \mathcal{Z}_\beta \left(\sigma,\mu\right)\ .
\end{gather}

\noindent Finally, in the $\beta\rightarrow\infty$ limit, the condition $\partial S/\partial\lambda=0$ yields

\begin{equation}
    \sum_{s=0}^\infty p_{\alpha q}(s)  \int \left\{ \mathrm{d}\hat{\pi} \right\}_s \left( \frac{\left\{\hat{h}\right\}_s}{\lambda - \left\{\hat{\omega}\right\}_s } \right)^2 = 1\ .
    \label{eq:IntegralCondition}
\end{equation}

\noindent A further simplification can be made by reducing the number of equations. This is done by inserting \eqref{eq:rho_hatSaddle} into \eqref{eq:rhoSaddle} to obtain

\begin{align}
     \rho(\sigma,\mu) = &\sum_{s=1}^\infty \frac{sp_{\alpha^{-1}q}(s)}{\alpha^{-1}q} \nonumber \\ &\times\int \left\{ \mathrm{d}\pi \right\}_{s-1} \Big\langle \delta\left( \sigma - \left( 1 - \sum_{\ell=1}^{s-1} \frac{K_\ell^2}{\omega_\ell} \right) \right) \delta\left( \mu - \sum_{\ell=1}^{s-1} \frac{K_\ell h_\ell}{\omega_\ell} \right) \Big\rangle_{\{K\}_{s-1}},
    \label{eq:rhoSaddle_Final}
\end{align}

\noindent where $\Big\langle\cdot\Big\rangle_{\{K\}_{s-1}}$ means averaging over $s-1$ random variables drawn from $p(K)$. Then, by substituting \eqref{eq:pi_hatSaddle} into \eqref{eq:piSaddle} we get

\begin{align}
   \pi\left(\omega, h\right) = &\sum_{s=1}^\infty\frac{sp_{\alpha q}(s)}{\alpha q}\nonumber \\ &\times \int \{\mathrm{d}\rho\}_{s-1} \Big\langle\delta\left(\omega-\left(\lambda-\sum_{\ell=1}^{s-1} \frac{K_\ell^2}{\sigma_\ell}\right)\right)\delta\left(h-\sum_{\ell=1}^{s-1}\frac{K_\ell \mu_\ell}{\sigma_\ell}\right) \Big\rangle_{\{K\}_{s-1}}.
    \label{eq:piSaddle_Final}
\end{align}

\noindent Furthermore, to express (\ref{eq:IntegralCondition}) in terms of $\rho$, we substitute (\ref{eq:pi_hatSaddle}) into  (\ref{eq:IntegralCondition}) and obtain

\begin{equation}
    \sum_{s=0}^\infty p_{\alpha q}(s)  \int \left\{ \mathrm{d}\rho \right\}_s \Big\langle\left( \frac{\sum_{\ell=1}^{s}\frac{K_\ell \mu_\ell}{\sigma_\ell}}{\lambda - \sum_{\ell=1}^s \frac{K^2_\ell}{\sigma_\ell} } \right)^2 \Big\rangle_{\{K\}_{s}}= 1\ .
    \label{eq:IntegralCondition_Final}
\end{equation}

\color{black}
\noindent One can, in principle, substitute \eqref{eq:rhoSaddle_Final} into \eqref{eq:piSaddle_Final} and \eqref{eq:IntegralCondition_Final}, and obtain self-contained equations for $\pi$, but this results in somewhat cumbersome expressions. \\

\subsection{The replicated partition function at the saddle point}
\color{black}
The final step in the analysis is to evaluate the saddle point form of the replicated partition function in the $\beta\rightarrow\infty$ limit. To this end, we use the saddle point forms of $\hat{\pi}$ and $\hat{c}$, [i.e. \eqref{eq:pi_hatSaddle} and \eqref{eq:c_hatSaddle} respectively] to obtain (arguments removed for ease of notation)

\begin{align}
     S_1 
    &\sim -\frac{nq\beta}{2} \int \mathrm{d}\pi \mathrm{d}\rho \Big\langle \frac{\left( h+ \frac{K\mu}{\sigma}\right)^2}{\omega-\frac{K^2}{\sigma}}-\frac{h^2}{\omega}\Big\rangle_K,
    \label{eq:S1Saddle}
\end{align}

\noindent where we used the definition of $\mathcal{Z}_\beta$ [Eq. \eqref{eq:Zbeta}] and evaluated the $\beta\rightarrow\infty$ asymptotic behaviour ($\sim$). Similarly,

\begin{equation}
     \tilde{S}_1  \sim -\frac{nq\beta}{2} \int \mathrm{d}\pi \mathrm{d}\rho \Big\langle \frac{\left( \mu+ \frac{Kh}{\omega}\right)^2}{\sigma-\frac{K^2}{\omega}}-\frac{\mu^2}{\sigma} \Big\rangle_K.
    \label{eq:S1_tildeSaddle}
\end{equation}
\noindent Next, we have

\begin{align}
    S_2 \sim \frac{nq\beta}{2} \int \mathrm{d}\hat{\pi} \sum_{s=0}^\infty \frac{s p_{\alpha q}(s)}{\alpha q} \int \{ \mathrm{d}\hat{\pi} \}_{s-1} \frac{\{\hat{h}\}_{s-1}+\hat{h}}{\lambda-\{\hat{\omega}\}_{s-1}-\hat{\omega}}\hat{h}\ .
\end{align}

\noindent Multiplying the last line by $1=\int \mathrm{d}\omega \mathrm{d}h  \delta\left(\omega-\left(\lambda-\{\hat{\omega}\}_{s-1}\right)\right)\delta\left(h-\{\hat{h}\}_{s-1}\right)$ and using the saddle point form of $\pi$ [Eq. \eqref{eq:piSaddle}], we have

\begin{align}
    S_2 &\sim \frac{nq\beta}{2}\int \mathrm{d}\hat{\pi} \mathrm{d}\pi \left(\frac{h+\hat{h}}{\omega-\hat{\omega}} \hat{h}\right)\ .\label{eq:S2_inter}
\end{align}

\noindent Then, by using the saddle point form of $\hat{\pi}$ [Eq. \eqref{eq:pi_hatSaddle}], we can further rewrite \eqref{eq:S2_inter} as 

\begin{equation}
    S_2 \sim \frac{nq\beta}{2} \int \mathrm{d}\pi \mathrm{d}\rho \Big\langle \frac{K\mu}{\sigma} \frac{h+\frac{K\mu}{\sigma}}{\omega-\frac{K^2}{\sigma}} \Big\rangle_K\ .
    \label{eq:S2Saddle}
\end{equation}

\noindent Following similar lines, we also conclude that

\begin{equation}
     \tilde{S}_2 \sim \frac{nq\beta}{2} \int \mathrm{d}\pi \mathrm{d}\rho \Big\langle \frac{Kh}{\omega} \frac{\mu+\frac{Kh}{\omega}}{\sigma-\frac{K^2}{\omega}} \Big\rangle_K.
     \label{eq:S2_tildeSaddle}
\end{equation}

\noindent \color{black} Lastly, considering the two equivalent forms of the interaction term [fifth line in Eq. \eqref{eq:SGauss} and \eqref{eq:SintDual}], its $\beta\rightarrow\infty$ limit can be written as

\begin{equation}
    S_{int} \sim \frac{nq\beta}{4} \int \mathrm{d}\pi \mathrm{d}\rho \Big\langle \frac{\left( h + \frac{K\mu}{\sigma} \right)^2}{\omega-\frac{K^2}{\sigma}}-\frac{h^2}{\omega} + \frac{\left( \mu + \frac{Kh}{\omega} \right)^2}{\sigma-\frac{K^2}{\omega}} -\frac{\mu^2}{\sigma} \Big\rangle_K\ .
    \label{eq:SintSaddle}
\end{equation}

\noindent Inserting \cref{eq:S1Saddle,eq:S1_tildeSaddle,eq:S2Saddle,eq:S2_tildeSaddle,eq:SintSaddle} into \eqref{eq:SGauss}, while noting that 

\begin{equation}
    \frac{K\mu}{\sigma}\frac{h+\frac{K\mu}{\sigma}}{\omega-\frac{K^2}{\sigma}} + \frac{Kh}{\omega}\frac{\mu+\frac{Kh}{\omega}}{\sigma-\frac{K^2}{\omega}} -\frac{1}{2}\left[ \frac{\left( h+ \frac{K\mu}{\sigma}\right)^2}{\omega-\frac{K^2}{\sigma}}-\frac{h^2}{\omega} +\frac{\left( \mu+ \frac{Kh}{\omega}\right)^2}{\sigma-\frac{K^2}{\omega}}-\frac{\mu^2}{\sigma}   \right] = 0\ ,
\end{equation}

\noindent the saddle point action eventually takes the form 

\begin{equation}
    S[\pi,\hat{\pi},\rho,\hat{\rho};\lambda] \sim  \frac{n\beta}{2\alpha}\lambda\ ,
    \label{eq:SSaddle}
\end{equation}
in the $n\to 0$ and $\beta\to\infty$ limits. Then, by inserting \eqref{eq:SSaddle} into \eqref{eq:ZnSaddleImplicit}, the replicated partition function at the saddle point becomes

\begin{equation}
    \Big\langle Z^n \Big\rangle \approx \mathrm{e}^{\frac{n\beta M}{2}\lambda}\ . 
    \label{eq:ZnSaddle}
\end{equation}

\noindent Finally, substituting \eqref{eq:ZnSaddle} into \eqref{eq:Lambda_Replica}, we obtain 

\begin{equation}
    \Big\langle \lambda_1 \Big\rangle = \lambda\ . \label{eq: Lambda Lambda}
\end{equation}

We recall at this point that the replica derivation started under the simplifying assumption that the $c_{ij}$'s are independent Bernouli random variables [See Eq. (\ref{eq:X distribution})]. This implies that the distribution of total number of nonzero elements in each row (column) - $p_{\alpha q}(s)$ ($p_{\alpha^{-1}q}(s)$) - naturally appearing in Eqs. \eqref{eq:rhoSaddle_Final} to \eqref{eq:IntegralCondition_Final}
is a Poisson distribution with unbounded support. However, due to \cite{Susca2019}, we know that these equations remain formally valid for any connectivity distribution $p(s)$. In our case, it is then necessary to consider the truncated Poisson distribution and manually amend the upper limit of the sums to account for the existence of a maximal number of nonzero elements in each row (column), $R$ ($C$). Putting everything together, in this section we have shown that by finding $\lambda$, $\pi$ and $\rho$ that solve the following system of recursive distributional equations supplemented by an integral constraint

\begin{gather}
     \pi\left(\omega, h\right) = \sum_{s=1}^R\frac{sp_{\alpha q}(s)}{\langle s\rangle _{p_{\alpha q}}} \int \{\mathrm{d}\rho\}_{s-1} \Big\langle\delta\left(\omega-\left(\lambda-\sum_{\ell=1}^{s-1} \frac{K_\ell^2}{\sigma_\ell}\right)\right)\delta\left(h-\sum_{\ell=1}^{s-1}\frac{K_\ell \mu_\ell}{\sigma_\ell}\right) \Big\rangle_{\{K\}_{s-1}} \nonumber \\   
    \rho(\sigma,\mu) = \sum_{s=1}^C \frac{sp_{\alpha^{-1}q}(s)}{\langle s\rangle _{p_{\alpha^{-1} q}}} \int \left\{ \mathrm{d}\pi \right\}_{s-1} \Big\langle \delta\left( \sigma - \left( 1 - \sum_{\ell=1}^{s-1} \frac{K_\ell^2}{\omega_\ell} \right) \right) \delta\left( \mu - \sum_{\ell=1}^{s-1} \frac{K_\ell h_\ell}{\omega_\ell} \right) \Big\rangle_{\{K\}_{s-1}}
    \nonumber \\
  \sum_{s=0}^R p_{\alpha q}(s)  \int \left\{ \mathrm{d}\rho \right\}_s \Big\langle\left( \frac{\sum_{\ell=1}^{s}\frac{K_\ell \mu_\ell}{\sigma_\ell}}{\lambda - \sum_{\ell=1}^s \frac{K^2_\ell}{\sigma_\ell} } \right)^2\Big\rangle_{\{K\}_{s}} = 1\,
    \label{eq:system}
\end{gather}

\noindent the typical largest eigenvalue of $\bm{J}$ is given by \eqref{eq: Lambda Lambda}. Note that for ease of notation, in \eqref{eq:system}, we used $p_{\alpha q}(s)$ ($p_{\alpha^{-1} q}(s)$) to denote the {\it truncated} Poisson distribution with parameter $\alpha q$ ($\alpha^{-1} q$), an upper cutoff $R$ ($C$), and $\langle s\rangle _{p_{\alpha q}}$ ($\langle s\rangle _{p_{\alpha^{-1} q}}$) denoting its average.

In Section \ref{PopDyn}, we will show that these integral equations can be efficiently solved using a Population Dynamics algorithm. In the next Section, we instead provide the theoretical framework to compute the probability density of the top eigenvector's components for diluted Wishart matrices. 
\color{black}

\section{Density of the top eigenvector's components}\label{sec:replica_eigenvector}

We now demonstrate how the results from the previous section can be applied to compute the average density of the top eigenvector's components for large $M,N$,
\begin{equation}
T\left(u\right)=\left\langle{\frac{1}{M}\sum_{i=1}^{M}\delta\left(u-v_1^{(i)}\right)}\right\rangle\ ,
\label{eq:instance_density}
\end{equation}
where once again, $\Big\langle\cdot\Big\rangle$ denotes averaging over different realisations of $\bm X$. We begin by outlining the strategy, highlighting its similarities and differences with the analysis in sec. \ref{Replica analysis}. We then carry it out to derive an expression for $T(u)$, which builds on the solution of Eq. \eqref{eq:system}.

To this end, we introduce the auxiliary partition function

\begin{equation}
Z^{(\beta)}_\epsilon(t,\bm{X};u)=\int\mathrm{d}\bm{v}\exp\left[\frac{\beta}{2}\left(\bm{v},\bm{J}\bm{v}\right)+\beta t\sum_{i}\delta_\epsilon\left(u-v_{i}\right)\right] \delta\left(\left|\bm{v}\right|^{2}-M\right)\ ,
\label{eq:Z Eigenvector Def}
\end{equation}
where $\delta_\epsilon$ is a smooth regulariser of the delta function and $\bm{J}=\bm{X}^T\bm{X}$. Due to the concentration of the Gibbs measure (see \eqref{canonicalpart}),

\begin{equation}
P_{\beta,\bm{X}}(\bm{v})=\frac{\exp\left(\frac{\beta}{2}\left(\bm{v},\bm{J}\bm{v}\right)\right) \delta\left(\left|\bm{v}\right|^{2}-M\right)}{\int\mathrm{d}\bm{v}'\exp\left( \frac{\beta}{2}(\bm{v}',\bm{J}\bm{v}')\right) \delta\left(\left|\bm{v}'\right|^{2}-M\right)}\ , 
\end{equation}
which localises around $\bm{J}$'s top eigenvector in the $\beta\to\infty$ limit, we can formally express $T(u)$ as

\begin{equation}
T(u)=\lim_{\beta\to\infty}\lim_{\epsilon\to 0^+}\frac{1}{\beta M}\frac{\partial}{\partial t}\left\langle\mathrm{Log}~Z^{(\beta)}_\epsilon(t,\bm{X};u)\right\rangle\Big|_{t=0}\ .
\label{eq:T(u) Formal}
\end{equation}
To evaluate $Z^{(\beta)}_\epsilon(t,\bm{X};u)$, we apply again the replica trick, leading to
\begin{equation}
T(u)=\lim_{\beta\to\infty}\lim_{\epsilon\to 0^+}\lim_{n\to 0}\frac{1}{\beta M}\frac{\partial}{\partial t}\frac{1}{n}\mathrm{Log}\left\langle [Z^{(\beta)}_\epsilon(t,\bm{X};u)]^n\right\rangle\Big|_{t=0}\
\ .\label{replicavecZeps}
\end{equation}
Since the structure of Eq. \eqref{eq:Z Eigenvector Def} resembles that of Eq. \eqref{canonicalpart}, with an additional $t$-dependent term, we expect that, in the large $N,M$ limit, the replicated partition function will once again take the form 
\begin{equation}
 \left\langle [Z^{(\beta)}_\epsilon(t,\bm{X};u)]^n\right\rangle\propto \int\mathcal{D}\varphi\mathcal{D}\hat{\varphi}\mathcal{D}\psi\mathcal{D}\hat{\psi}\mathrm{d}\vec{\lambda}\exp\left \{\sqrt{NM}S^{(\beta)}_{n}\left[\phi,\hat{\phi},\psi, \hat{\psi}, \vec{\lambda};t,\epsilon;u\right]\right\}\ .
 \label{eq:Zn Functional Integral Eigenvecor}
\end{equation}
The above structure will then enable us to employ the replica-symmetric ansatz [i.e. represent the fields as a superposition of uncountably infinite Gaussians, see Eqs. \eqref{eq:phiGauss}-\eqref{eq:Zbeta}]
 and perform a saddle-point evaluation for large $N,M$
\begin{equation}
 \left\langle [Z^{(\beta)}_\epsilon(t,\bm{X};u)]^n\right\rangle\approx \exp\left \{\sqrt{NM}S^{(\beta)}_{n}\left[\pi^\star,\hat{\pi}^\star,\rho^\star, \hat{\rho}^\star, \vec{\lambda}^\star;t,\epsilon;u\right]\right\}\ ,\label{actionvec}
\end{equation}
where the starred objects represent the saddle point forms of $\pi,\hat{\pi},\rho,\hat{\rho},\vec\lambda$, found through the corresponding stationary conditions. Since the partial derivative $\frac{\partial}{\partial t}$ in \eqref{eq:T(u) Formal} only acts on terms containing any explicit dependence on $t$, and not through any other indirect functional dependence, $t$ can be safely set to zero in the resulting saddle-point equations. Consequently, $\pi^\star,\hat{\pi}^\star,\rho^\star, \hat{\rho}^\star, \vec{\lambda}^\star$ satisfy the same saddle-point equations derived in sec. \ref{Saddle Point analysis}.

Inserting \eqref{actionvec} into \eqref{replicavecZeps} and assuming that the leading $n\to 0$ behaviour of the action at the saddle point is given by
\begin{equation}
S^{(\beta)}_{n}\left[\pi^\star,\hat{\pi}^\star,\rho^\star, \hat{\rho}^\star, \vec{\lambda}^\star;t,\epsilon;u\right]\sim n s_\beta\left(t,\epsilon;u\right) +o(n)\ ,\label{sbeta}
\end{equation}
the final expression for the average density of top eigenvector's components is obtained by inserting Eqs. \eqref{replicavecZeps}, \eqref{actionvec} and \eqref{sbeta} into \eqref{eq:T(u) Formal}
\begin{equation}
T(u)=\lim_{\beta\to\infty}\frac{\alpha}{\beta}s_\beta'\left(0,0;u\right)\ ,\label{betalimitfinal}
\end{equation}
where $(\cdot)'$ stands for differentiation with respect to $t$. Since the saddle-point equations for $\pi, \hat{\pi}, \rho, \hat{\rho}, \vec{\lambda}$ are identical to those derived in sec. \ref{Saddle Point analysis}, the remaining challenge is to identify $s_\beta(t, \epsilon; u)$ and evaluate \eqref{betalimitfinal}. \\

To this end, we apply this strategy to our matrix $\bm J=\bm X^T\bm X\ $, where $\bm X$'s entries follow the distribution given in Eq. \eqref{eq:X Def}. Exponentiating the replicated partition function by following the same lines as in sec. \ref{Exponentiating}, we obtain

\begin{align}
    & \left\langle [Z^{(\beta)}_\epsilon(t,\bm{X};u)]^n\right\rangle \propto \int \left( \prod_{a=1}^n \mathrm{d}\bm{v}_a \mathrm{d}\bm{u}_a \mathrm{d}\lambda_a\right) \exp\left(-\frac{\beta}{2}\sum_{a=1}^n \sum_{j=1}^N u_{ja}^2\right) \exp\left(\mathrm{i}M \frac{\beta}{2}\sum_{a=1}^n \lambda_a\right)  \nonumber \\
    & \times \exp\left(-\mathrm{i} \frac{\beta}{2}\sum_{a=1}^n\sum_{i=1}^M \lambda_a v_{ia}^2\right) \exp\left\{ \frac{q}{\sqrt{NM}}\sum_{i=1}^M\sum_{j=1}^N \left[\Big\langle\exp\left(\beta K\sum_{a=1}^n v_{ia}u_{ja}\right)\Big\rangle_K -1\right] \right\} \nonumber \\
    &\times\exp\left[\beta t\sum_{a=1}^n\sum_{i=1}^M\delta_\epsilon\left(u-v_{ia}\right)\right]\ .
    \label{eq:ZnEigvec}
\end{align}
Comparing this expression with Eq. \eqref{eq:ZnExplic}, it is natural to define the same functional order parameters as in Eq. \eqref{eq:RepDensPsi},

\begin{gather}
    \phi(\Vec{v})=\frac{1}{M} \sum_{i=1}^M \prod_{a=1}^n \delta\left( v_a - v_{ia} \right) \label{eq:RepDensPhi1}\ , \\
    \psi(\Vec{u})=\frac{1}{N} \sum_{j=1}^N \prod_{a=1}^n \delta\left( u_a - u_{ja} \right)\ .
    \label{eq:RepDensPsi1}
\end{gather}
Then, by following the same lines as in sec. \ref{Functional Integral Representation}, we see that the functional-integral form of \eqref{eq:ZnEigvec} is identical to the one in \eqref{eq:ZnPathIntegral}, except for the term $S_2$. Hence, the functional integral representation of the replicated partition function can indeed be expressed as 
\begin{equation}
 \left\langle [Z^{(\beta)}_\epsilon(t,\bm{X};u)]^n\right\rangle\propto \int\mathcal{D}\varphi\mathcal{D}\hat{\varphi}\mathcal{D}\psi\mathcal{D}\hat{\psi}\mathrm{d}\vec{\lambda}\exp\left \{\sqrt{NM}S^{(\beta)}_{n}\left[\phi,\hat{\phi},\psi, \hat{\psi}, \vec{\lambda};t,\epsilon;u\right]\right\}\ ,
 \label{eq:Zn Functional Integral Eigenvecor1}
\end{equation}
with the action given by
\begin{align}
S^{(\beta)}_{n}\left[\phi,\hat{\phi},\psi, \hat{\psi}, \vec{\lambda};t,\epsilon;u\right] &=S_{1}\left[\phi,\hat{\phi}\right]+S_{2}\left[\hat{\phi}, \vec\lambda; t; \epsilon;u\right]\nonumber \\
&+\tilde{S}_{1}\left[\psi,\hat{\psi}\right]+\tilde{S}_2\left[\hat{\psi}\right]+S_{3}\left[\vec\lambda\right]+S_{int}\left[\phi,\psi\right]\ ,
\end{align}
where all contributions other than $S_2$ are identical to those defined in Eqs. \eqref{eq:S1}-\eqref{eq:Sint}, and the $t$ and $\epsilon$ dependence is confined to $S_2$, which is now given by
\begin{equation}
S_{2}\left[\hat{\phi},\vec{\lambda};t,\epsilon;u\right] =  \frac{1}{\alpha}\mathrm{Log}\int\mathrm{d}\vec{v}\exp\left[ -\mathrm{i}\frac{\beta}{2}\sum_{a}^n\lambda_{a}v_{a}^{2}+\beta t\sum_{a}^n\delta_\epsilon\left(u-v_{a}\right)+\mathrm{i}\hat{\phi}\left(\vec{v}\right)\right]\ .\label{S4mu}
\end{equation}
We then follow the same strategy as in sec. \ref{replicasymmetricansatz}, and enforce the replica symmetric ansatz by representing the functional order parameters as a superposition of uncountably infinite Gaussians [see Eqs. \eqref{eq:phiGauss}-\eqref{eq:Zbeta}]. Specifically, we recall that
\begin{equation}
    \mathrm{i}\hat{\phi}\left(\vec{v}\right) = \hat{c}\int \mathrm{d}\hat{\pi} \prod_{a=1}^n \mathrm{e}^{\frac{\beta}{2}\hat{\omega} v_a^2 + \beta \hat{h} v_a}\ .
\end{equation}
Substituting this representation into Eq. \eqref{S4mu} yields the following leading $n\to 0$ behaviour,
\begin{align}
\nonumber S_{2} &\simeq  \frac{\hat{c}}{\alpha}+\frac{n}{\alpha}\sum_{s=0}^{\infty}p_{\hat{c}}\left(s\right)\int\{ \mathrm{d}\hat{\pi}\} _{s}~\mathrm{Log}\int\mathrm{d}v\exp\left[ -\mathrm{i}\frac{\beta}{2}\lambda v^{2}\right.\\
&\left.+\beta t\delta_\epsilon\left(u-v\right)+\frac{\beta}{2}\{\hat{\omega}\}_s v^{2}+\beta\{\hat{h}\}_s v\right]\ .
\end{align}
Therefore, we can identify the function $s_\beta(t,\epsilon;u)$ in \eqref{sbeta} as
\begin{align}
\nonumber s_\beta(t,\epsilon;u) &=  \frac{1}{\alpha}\sum_{s=0}^{\infty}p_{\hat{c}}\left(s\right)\int\{ \mathrm{d}\hat{\pi}\} _{s}~\mathrm{Log}\int\mathrm{d}v\exp\left[ -\frac{\beta}{2}\lambda v^{2}+\beta t\delta_\epsilon\left(u-v\right)
\right.\\
&\left.+ \frac{\beta}{2}\{\hat{\omega}\}_s v^{2}+\beta\{\hat{h}\}_s v\right]\ ,
\end{align}
with $\mathrm{i}\lambda\equiv\lambda$ solving \eqref{eq:system} as before. Taking the $t$-derivative and setting $t$ and $\epsilon$ to zero, while recalling that $\hat{c}=\alpha q$, we obtain
\begin{align}
\nonumber s^\prime_\beta(0,0;u) &= \frac{\beta}{\alpha}\sum_{s=0}^{\infty}p_{\alpha q}\left(s\right)\int\{ \mathrm{d}\hat{\pi}\} _{s}\frac{\exp\left[ -\frac{\beta}{2}(\lambda-\{\hat{\omega}\}_s) u^{2}+\beta\{\hat{h}\}_s u\right]}{\int\mathrm{d}v\exp\left[ -\frac{\beta}{2}(\lambda-\{\hat{\omega}\}_s)v^{2}+\beta\{\hat{h}\}_s v\right]}\ .\end{align}
Taking the $\beta\to\infty$ limit and inserting the result into Eq. \eqref{betalimitfinal}, we find that
\begin{equation}
T(u)  =\sum_{s=0}^{\infty}p_{\alpha q}(s)\int\left\{\mathrm{d}\hat{\pi}\right\} _{s}\delta\left(u-\frac{\{ \hat{h}\} _{s}}{\lambda-\{ \hat{\omega}\} _{s}}\right)\ .\label{eq:density_vector_general}
\end{equation}
Finally, using the saddle point form of $\hat{\pi}$ [Eq. \eqref{eq:c_hatSaddle}] to express it via $\rho$, and truncating the Poisson distribution as before, we obtain
\begin{equation}
T(u)  =\sum_{s=0}^{R}p_{\alpha q}(s)\int\left\{\mathrm{d}\rho\right\} _{s}\left\langle\delta\left(u-\frac{\sum_{\ell=1}^s\frac{K_\ell \mu_\ell}{\sigma_\ell}}{\lambda-\sum_{\ell=1}^s \frac{K^2}{\sigma_\ell}}\right)\right\rangle_{\{K\}_s}\ .\label{eq:density_vector_general1}
\end{equation}
Putting everything together, after solving Eq. \eqref{eq:system} for $\rho$ and $\lambda$, these can then be used to sample the integral in \eqref{eq:density_vector_general1} and obtain the density of the top eigenvector's components. The algorithmic way to do this is explained in the next Section. \\

\color{black}
\section{Population dynamics}
\label{PopDyn}
\begin{figure}
\centering
    \begin{subfigure}{0.45\textwidth}
        \centering
        \includegraphics[width=\linewidth]{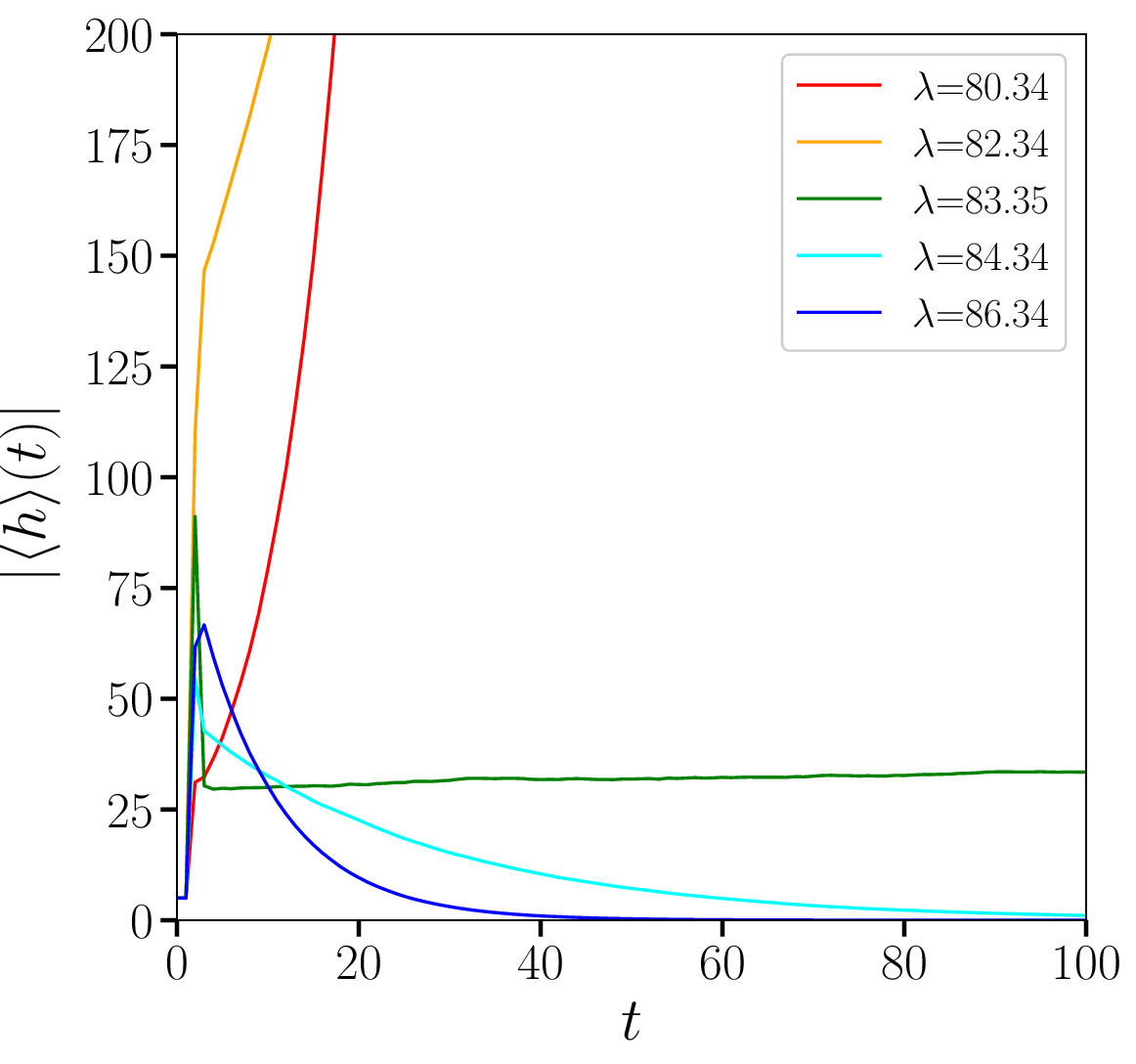}
        \label{fig:gap_40_300}
    \end{subfigure}
    \hfill
    \begin{subfigure}{0.45\textwidth}
        \centering
        \includegraphics[width=\linewidth]{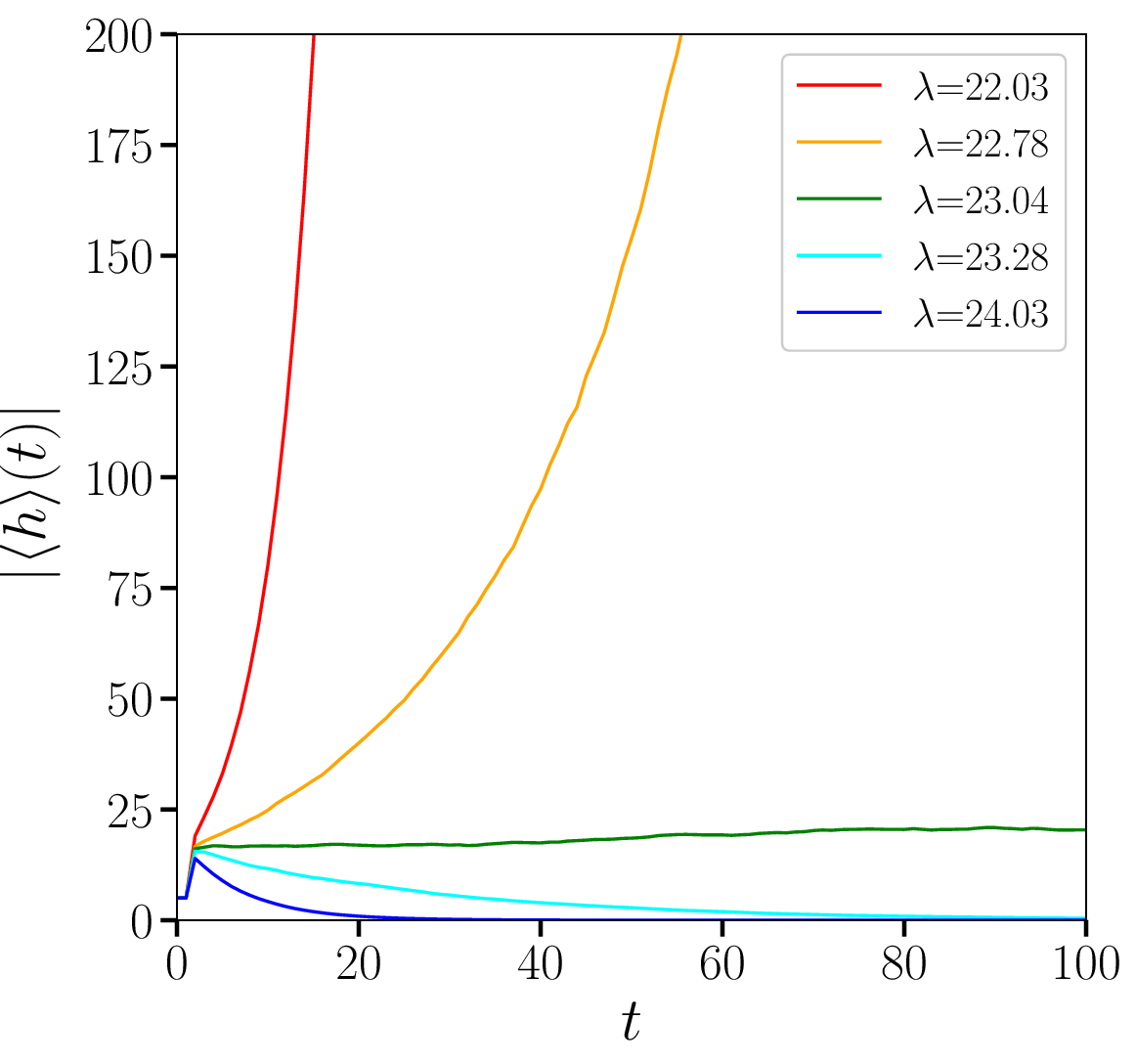}
        \label{}
    \end{subfigure}
\caption{Evolution of the first moment of the $h$ population in absolute value, $|\langle h\rangle|(t)$, according to the population dynamics algorithm as outlined in section \ref{PopDyn}, with population size of $N_P=10^5$ and where $t$ is measured in sweeps. The control parameters in this figure are chosen as $q=8$, $\alpha=\sqrt{5/4}$ and $p(K)=\delta_{K,1}$ for the left figure and $p(K)=\Theta(1-K)\Theta(K)$ for the right figure. In both figures the maximal number of nonzero elements in each row is set to $R=70$ and in each column to $C=60$. The target error tolerance was set to $\Delta=0.1$. \color{black} The different curves correspond to ascending values of $\lambda$ (top to bottom), the parameter that governs the convergence of the algorithm, which was initialised as $\lambda_{\rm initial}=250$ in both figures. \color{black} For $\lambda<\Big\langle\lambda_1\Big\rangle$ (red and orange lines) the population diverges, for $\lambda>\Big\langle\lambda_1\Big\rangle$ (blue and cyan lines) it vanishes, and only when $\lambda=\Big\langle\lambda_1\Big\rangle$ (green line), stability is reached. The rate of divergence/decay depends on the amount by which $\lambda$ deviates from $\Big\langle\lambda_1\Big\rangle$. }
\label{fig:PopDyn}
\end{figure}

In this section we briefly present the population dynamics algorithm \cite{Kuhn2007,Mezard2000,Zdeborov2016}, which can be used to numerically solve the system given by \eqref{eq:system}. Different incarnations of this algorithm have been used in a number of problems recently \cite{Susca2019,Susca2020,Susca2021,Kuhn2017,Kuhn2024,Bartolucci2024}. 
\color{black}
For a specified set of inputs ${q, \alpha, p(K),R,C}$ \color{black} and a target error tolerance $\Delta$, the algorithm outputs the theoretical value of $\langle \lambda_1 \rangle$, with an uncertainty $\pm\Delta /2$:

\begin{enumerate}

    \item Initialise the real parameter $\lambda$ to a ``large'' value (using the estimate in \ref{Upper Bound}).

    \item Randomly initialise two sets of coupled populations, each of size $N_P$, $\{\left(\omega_i,h_i\right)\}_{1\leq i\leq N_P}$ and $\left\{\left(\sigma_i,\mu_i\right)\right\}_{1\leq i\leq N_P}$.
    
    \item Generate a random $s\sim\frac{sp_{\alpha q}\left(s\right)}{\langle s\rangle}$, where $p_{\alpha q}(s)$ is a truncated Poisson distribution with parameter $\alpha q$ and upper cutoff $R$.

    \item Draw $s-1$ i.i.d. random variables $K_\ell$ from $p(K)$.

    \item Select $s-1$ random pairs $\{(\sigma_\ell,\mu_\ell)\}_{\ell=1}^{s-1}$ from the population, compute

    \begin{gather}
        \omega^{(\text{new})} = \lambda -\sum_{\ell=1}^{s-1} \frac{K_\ell^2}{\sigma_\ell}\ ,\\
        h^{(\text{new})} = \sum_{\ell=1}^{s-1} \frac{K_\ell \mu_\ell}{\sigma_\ell}\ ,
    \end{gather}

    and replace a randomly selected pair $(\omega_r,h_r)$ with $(\omega^{(\text{new})},h^{(\text{new})})$.

    \item Generate a random $s\sim \frac{sp_{\alpha^{-1}q\left(s\right)}}{\langle s\rangle}$, where $p_{\alpha^{-1} q}(s)$ is a truncated Poisson distribution with parameter $\alpha^{-1} q$ and upper cutoff $C$.

    \item Draw $s-1$ i.i.d. random variables $K_\ell$ from $p(K)$.
    
    \item Select $s-1$ random pairs $\{(\omega_\ell,h_\ell)\}_{\ell=1}^{s-1}$ from the population, compute
    
    \begin{gather}
        \sigma^{(\text{new})} = 1 -\sum_{\ell=1}^{s-1} \frac{K_\ell^2}{\omega_\ell}\ ,\\
        \mu^{(\text{new})} = \sum_{\ell=1}^{s-1} \frac{K_\ell h_\ell}{\omega_\ell}\ ,
    \end{gather}

    and replace a randomly selected pair $(\sigma_r,\mu_r)$ with $(\sigma^{(\text{new})},\mu^{(\text{new})})$.
    \color{black}
    \item After every sweep, monitor the populations' first moment.
    \begin{itemize}
        \item If any one of them shrinks to zero. Set $\lambda^{(\text{new})} = \lambda - \Delta$ and return to (ii).
        \item If any one of them explodes, set $\Big\langle\lambda_1\Big\rangle = \lambda + \Delta/2$ and exit the algorithm.
    \end{itemize}
    
    \item Return to (iii).
\end{enumerate}
\color{black}
The nature of the algorithm ensures that the only value of the (real) parameter $\lambda$ under which stability is reached is the one corresponding to $\Big\langle\lambda_1\Big\rangle$ \cite{Susca2019}. When $\lambda<\Big\langle\lambda_1\Big\rangle$ the $h$ and $\mu$ populations \color{black} will explode, and for $\lambda>\Big\langle\lambda_1\Big\rangle$ they will shrink to zero. Consequently, one can monitor the populations' stability by examining the time-evolution of their first moment, as shown in Fig. \ref{fig:PopDyn}. Another observation is that the rates at which the populations diverge and vanish increase as the value of $\lambda$ deviates from $\Big\langle\lambda_1\Big\rangle$. Furthermore, the stable regime is highly peaked around $\lambda=\Big\langle\lambda_1\Big\rangle$, which allows us to pinpoint the value of $\Big\langle\lambda_1\Big\rangle$ with very high precision. 

Specifying to the case where $\lambda = \langle \lambda_1 \rangle$ and nontrivial stability is achievable, it is possible to identify multiple fixed points for the densities $\pi$ and $\rho$ \color{black} that satisfy the first two equations in \eqref{eq:system} by adjusting the initial populations. However, incorporating the third equation in \eqref{eq:system} uniquely determines the solution. Once the algorithm identifies the value of $\lambda$ that allows nontrivial stable populations, the third condition in \eqref{eq:system} can be fulfilled by rescaling the $h$ and $\mu$ populations\color{black}, yielding a solution that satisfies the full set of equations \eqref{eq:system} in its entirety. This rescaling is always allowed due to the linear nature of the recursion governing their updates \cite{Susca2019}.

Given the behaviour described above, the strategy for pinning down the value of $\lambda$ under which stability can be reached, is to start with a large value, determined by a proper upper bound for $\Big\langle\lambda_1\Big\rangle$. Then, while running the algorithm, one monitors the time evolution of the populations' first moment, and gradually decreases the value of $\lambda$ until they stabilise. A plausible upper bound that can be used as a starting point is $\lambda^\star=\left[\max\left(|\zeta^-|,|\zeta^+|\right)\right]^2RC$, where $\zeta^-$ ($\zeta^+$) is the lower (upper) bound of the support of $p(K)$, while $R$ and $C$ are the maximal numbers of nonzero elements in each row and in each column respectively \color{black} (See \ref{Upper Bound} for a proof). Once the populations stabilise and the typical largest eigenvalue is determined, one can use them to obtain the density of the top eigenvector's components via \eqref{eq:density_vector_general1}.

\begin{figure}
\centering
\includegraphics[width=0.45\linewidth]{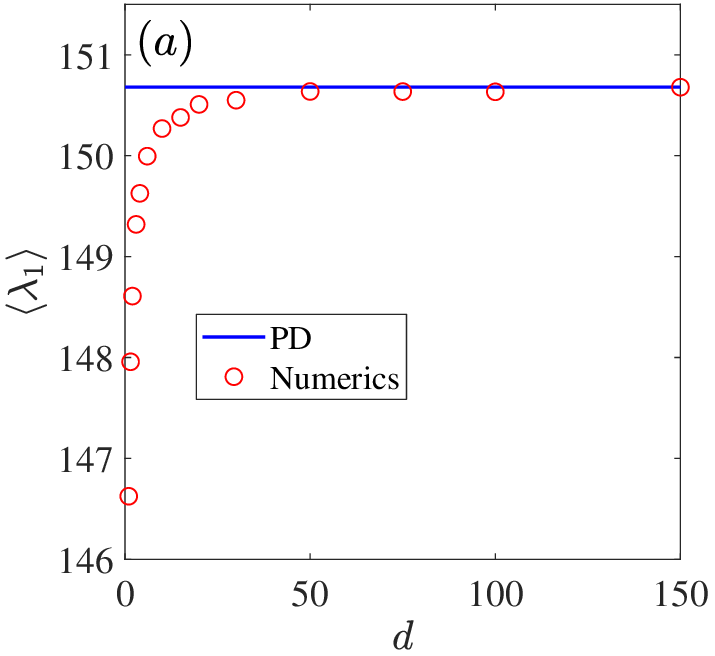} \hspace{1.2cm}
\includegraphics[width=0.45\linewidth]{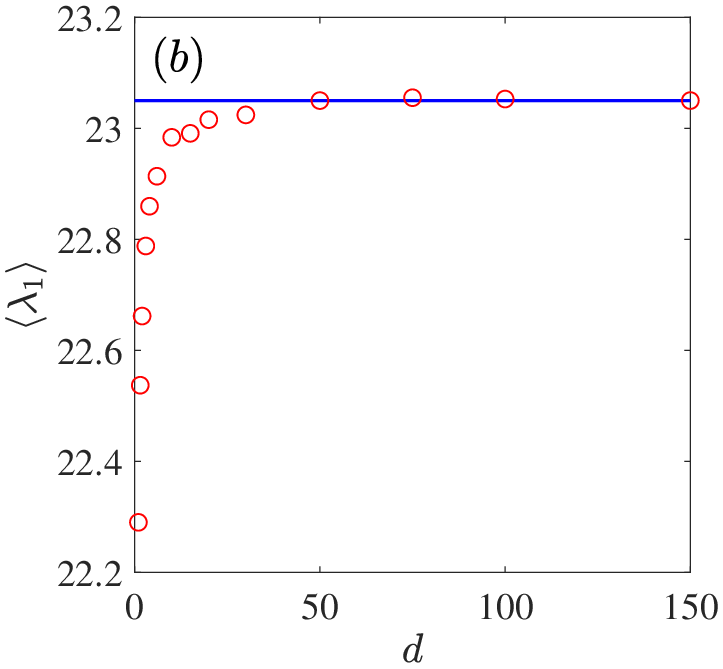}
\caption{Scaling of $\Big\langle\lambda_1\Big\rangle$ with the dimensions of the matrix $\bm{X}$. This figure shows $\Big\langle\lambda_1\Big\rangle$, collected from direct numerical diagonalisation of $10^2$ realisations of $\bm{J}$ (circles), as the size of the matrix $\bm{X}$ is increased, while the ratio $\alpha=\sqrt{N/M}$ is kept fixed. The scaling parameter $d$ is defined such that each data point was obtained using a matrix $\bm{X}$ of size $(100\cdot d)\times(80\cdot d)$. The solid blue line represents the results obtained from the replica analysis, using the population dynamics algorithm, using populations of size $N_P=10^5$. The set of control parameters used here is (a) $\alpha=\sqrt{5/4}$, $q=11.8$ and $p(K)=\delta_{K,1}$; (b) $\alpha=\sqrt{5/4}$, $q=8$ and $p(K)=\Theta(K)\Theta(1-K)$. In both figures the maximal number of nonzero elements in each row is set to $R=70$ and in each column to $C=60$. The target error tolerance was set to $\Delta=0.1$. \color{black} As can be observed from the figure, even for a relatively small matrix of size $100\times 80$, finite size effects are responsible for a deviation of only up to $\sim 4\%$ from the analytical result. For a matrix $\sim 50$ times bigger than that, this deviation drops below the measurement's resolution.
}
\label{fig:Scaling}
\end{figure}

\color{black} In Fig. \ref{fig:Scaling} we present the scaling of $\Big\langle\lambda_1\Big\rangle$ with the dimensions of the matrix $\bm{X}$, under the following choice of control parameters: (a) $\alpha=\sqrt{5/4}$, $q=11.8$ and $p(K)=\delta_{K,1}$; (b) $\alpha=\sqrt{5/4}$, $q=8$ and $p(K)=\Theta(K)\Theta(1-K)$, with $\Theta(\cdot)$ being the Heaviside function [i.e. $K\in (0,1)$ with uniform probability]. In both figures the maximal number of nonzero elements in each row is set to $R=70$ and in each column to $C=60$. The target error tolerance was set to $\Delta=0.1$. \color{black} As outlined in section \ref{Replica analysis}, we computed the leading behaviour of $\Big\langle\lambda_1\Big\rangle$ as both of $\bm{J}$'s linear dimensions tend to infinity. Therefore, our analysis does not account for any finite size effects. However, in Fig. \ref{fig:Scaling} we show that these corrections are negligible compared to the leading behaviour, which is perfectly captured by our analysis. Specifically, even for a relatively small matrix of size $100\times 80$, finite size corrections are responsible for a deviation of merely $\sim 4\%$. When the matrix size is further increased, the numerical results quickly align with our analytical results, to the extent that the two are indistinguishable within our measurement's resolution. 

\begin{figure}
\centering
\includegraphics[width=0.45\linewidth]{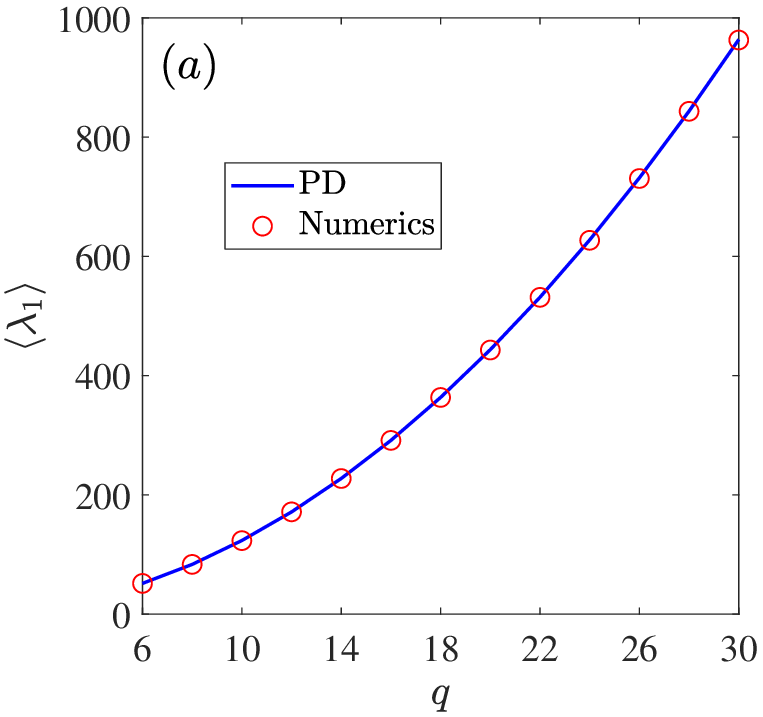} \hspace{1.2cm}
\includegraphics[width=0.45\linewidth]{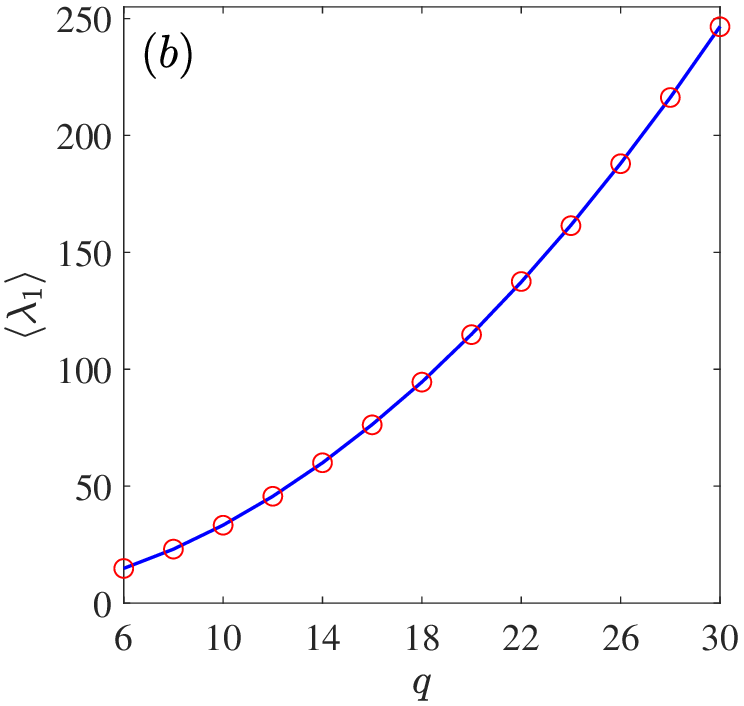}
\caption{We show $\Big\langle\lambda_1\Big\rangle$ as obtained by both population dynamics (solid line) and direct numerical diagonalisation (circles) as a function of $q$, which regulates the average density of nonzero elements in $\bm{X}$. For this analysis, we used $\alpha = \sqrt{5/4}$ and set the weight distribution to (a) $p(K) = \delta_{K,1}$, and (b) $p(K)=\Theta(1-K)\Theta(K)$. In both figures the maximal number of nonzero elements in each row is set to $R=70$ and in each column to $C=60$. The target error tolerance was set to $\Delta=0.1$. \color{black} In view of Fig. \ref{fig:Scaling}, the numerical data represents an average over $10^2$ realisations of $\bm{X}$, each of fixed and large dimensions $5,000\times 4,000$. Under these conditions, the numerical and analytical results are in very strong  agreement.}
\label{fig:ReplicaVsNumerics}
\end{figure}

In Fig. \ref{fig:ReplicaVsNumerics} we compare results for $\Big\langle\lambda_1\Big\rangle$ obtained from the replica analysis (solid line) and direct numerical diagonalisation (circles), as a function of $q$, which regulates the average density of nonzero elements in $\bm{X}$. In this figure we chose $\alpha=\sqrt{5/4}$ and the weight distributions (a) $p(K)=\delta_{K,1}$; (b)  $p(K)=\Theta(1-K)\Theta(K)$. In both figures the maximal number of nonzero elements in each row is set to $R=70$ and in each column to $C=60$. The target error tolerance was set to $\Delta=0.1$. \color{black} In light of Fig. \ref{fig:Scaling}, the numerical data was obtained by averaging over $10^2$ realisations of $\bm{X}$ with a fixed size of $5,000\times4,000$, such that finite size corrections are negligible. Within this framework, we find excellent agreement between the numerical and analytical results.

\begin{figure}
\centering
\includegraphics[width=0.45\linewidth]{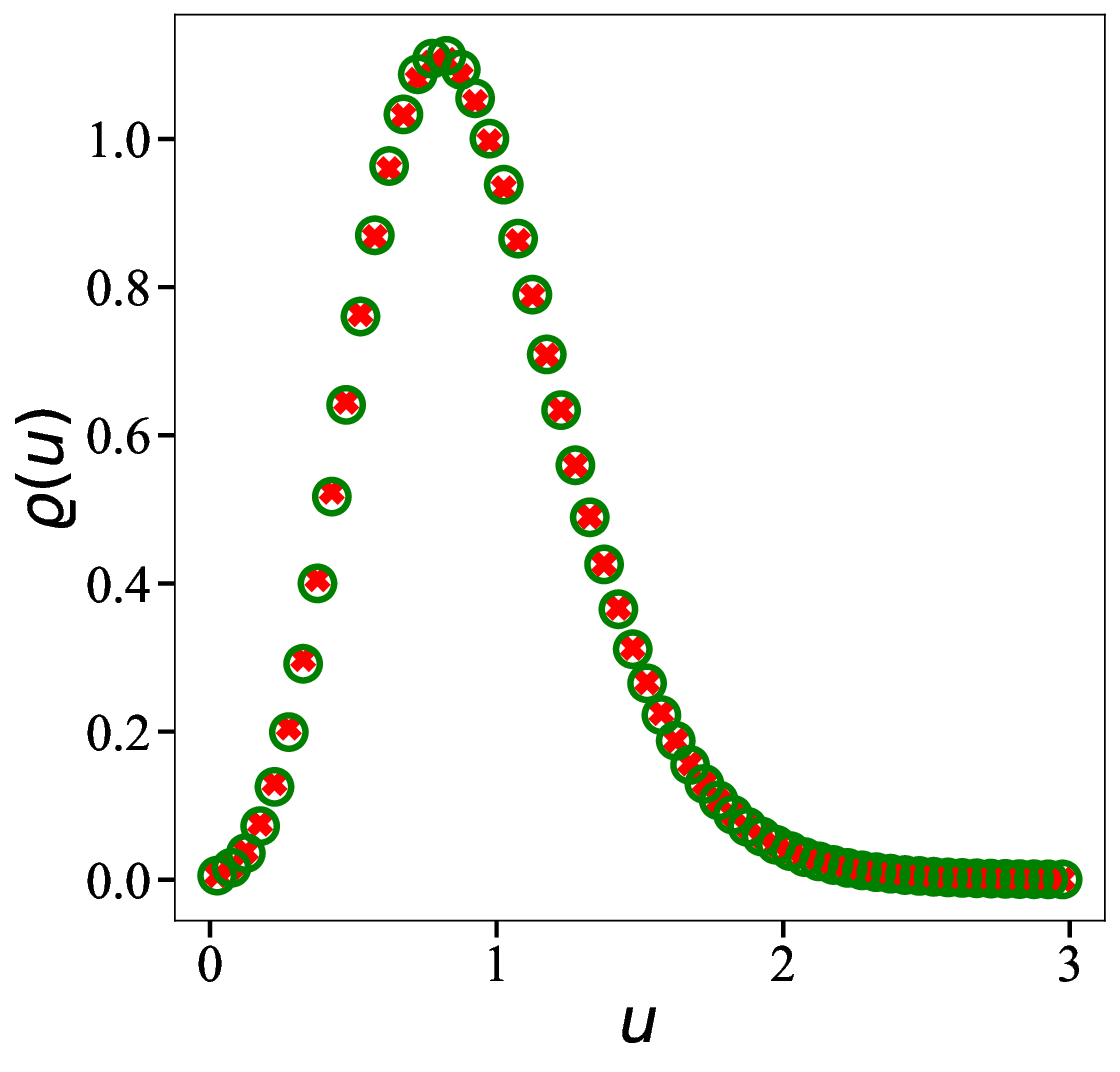} \hspace{1.2cm}
\includegraphics[width=0.45\linewidth]{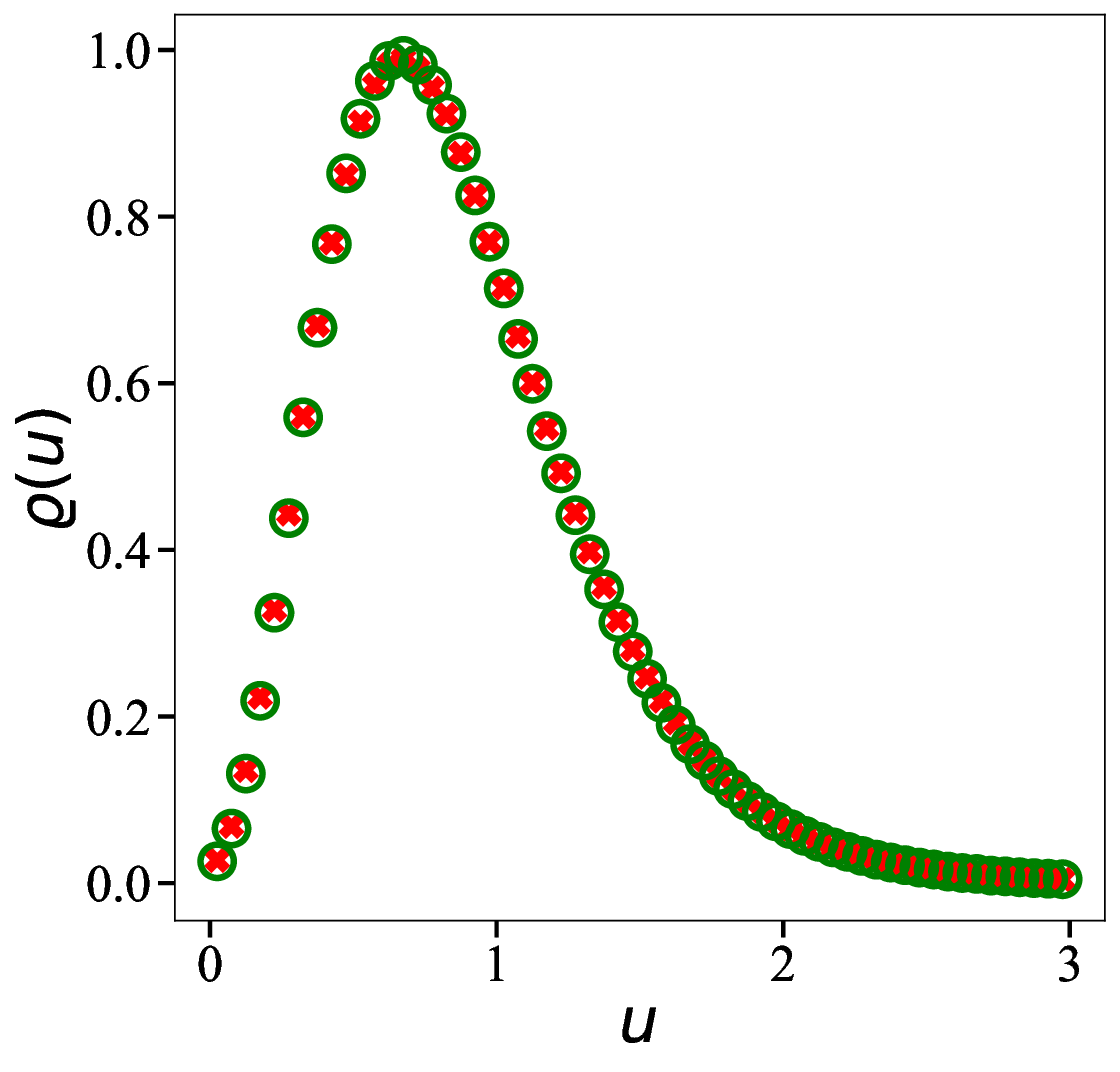}
\caption{We compare the results for $T(u)$, obtained from \eqref{eq:density_vector_general1} (red crosses) and direct numerical diagonalisation (green circles).  For this analysis, we used $\alpha = \sqrt{5/4}$ and set the weight distribution to (a) $p(K) = \delta_{K,1}$, and (b) $p(K)=\Theta(1-K)\Theta(K)$ as in Fig. (\ref{fig:ReplicaVsNumerics}). In both figures the maximal number of nonzero elements in each row is set to $R=70$ and in each column to $C=60$. The target error tolerance was set to $\Delta=0.1$. \color{black} The numerical data represents an average over $10^2$ realisations of $\bm{X}$, each of fixed and large dimensions $5,000\times 4,000$. The numerical and analytical results are in very strong  agreement. The resolution of the density, $\Delta u$, has been set at $\Delta u=0.05$.\color{black}
}
\label{fig:Eigenvector Components}
\end{figure}

Building on the results for $\langle \lambda_1\rangle$, in Fig. \ref{fig:Eigenvector Components}, we compare the results for $T(u)$, obtained from Eq. \eqref{eq:density_vector_general1} (red crosses) and direct numerical diagonalisation (green circles).  For this analysis, we used the same settings as in Fig. (\ref{fig:ReplicaVsNumerics}). The numerical and analytical results are again in very strong  agreement.

The density $T(u)$ was numerically evaluated using a procedure based on a population generated by the algorithm outlined in Section \ref{PopDyn}. We initially choose a resolution for our density, denoted by $\Delta u$, and split up the interval $[0,3]$ into bins of size $\Delta u$. We then generate a stable population following the algorithm of Section \ref{PopDyn} and randomly sample members of the population in order to evaluate the value 
\begin{equation}
\frac{\sum_{\ell=1}^s\frac{K_\ell \mu_\ell}{\sigma_\ell}}{\lambda-\sum_{\ell=1}^s \frac{K^2}{\sigma_\ell}} \ . \label{Algorithm Value}
\end{equation}
Each time the computed value of \eqref{Algorithm Value} fell within a given bin, a count of one was added to that bin. This procedure was performed many times, after which the bin counts were normalised in order to produce the numerical density $T(u)$.

\section{The dense limit}
\label{Large q}

Below, we demonstrate how taking the large $q$ limit recovers the familiar results of the noncentral Wishart ensemble, by following the same lines as in Ref. \cite{Susca2019}. To ensure a finite largest eigenvalue in this limit, we rescale the bond weights as
\begin{equation}
    K_{ij} = \frac{\tilde{K}_{ij}}{q}\ ,
    \label{eq:K Scaling}
\end{equation}
and assume that $\langle\tilde{K}\rangle$ and $\langle\tilde{K}^2\rangle$ are nonzero and of $\mathcal{O}(1)$. Note that this scaling differs from that used in Ref.~\cite{Susca2019}, which is $K=\tilde{K}/\sqrt{q}$, due to a subtle but important difference in the underlying assumptions. In ~\cite{Susca2019}, the authors consider a sparse \textit{central} model, in which the nonzero entries satisfy $\langle K \rangle = 0$. In the dense limit, they recover the upper edge of the semicircle law. In contrast, our model is \textit{noncentral}, in the sense that $\langle K \rangle \ne 0$, which causes the largest eigenvalue to detach from the bulk of the spectrum. As we will later demonstrate, our scaling ensures that the detached $\langle \lambda_1 \rangle$ remains of $\mathcal{O}(1)$, whereas alternative scalings would yield a vanishing or diverging result in the $q \to \infty$ limit. Inserting \eqref{eq:K Scaling} into the first two lines of \eqref{eq:system}, we obtain

\begin{align}
     \pi(\omega,h) = &\sum_{s=1}^\infty \frac{sp_{\alpha q}(s)}{\alpha q} \int \left\{ \mathrm{d}\rho \right\}_{s-1} \Big\langle \delta\left( \omega - \left( \lambda - \frac{1}{(\alpha q)^2}\sum_{\ell=1}^{s-1} \frac{\alpha^2\tilde{K}_\ell^2}{\sigma_\ell} \right) \right) \nonumber \\
     &\times \delta\left( h - \frac{1}{\alpha q}\sum_{\ell=1}^{s-1} \frac{\alpha\tilde{K}_\ell \mu_\ell}{\sigma_\ell} \right) \Big\rangle_{\{\tilde{K}\}_{s-1}}
     \label{eq:Large c pi}
\end{align}

\noindent and

\begin{align}
    \rho(\sigma,\mu) = &\sum_{s=1}^\infty \frac{sp_{\alpha^{-1} q}(s)}{\alpha^{-1} q}\nonumber  \int \left\{ \mathrm{d}\pi \right\}_{s-1} \Big\langle \delta\left( \sigma - \left( 1 - \frac{1}{(\alpha^{-1} q)^2}\sum_{\ell=1}^{s-1} \frac{\alpha^{-2}\tilde{K}_\ell^2}{\omega_\ell} \right) \right) \nonumber \\
    &\times \delta\left( \mu - \frac{1}{\alpha^{-1} q}\sum_{\ell=1}^{s-1} \frac{\alpha^{-1}\tilde{K}_\ell h_\ell}{\omega_\ell} \right) \Big\rangle_{\{\tilde{K}\}_{s-1}}\ .
    \label{eq:Large c rho}
\end{align}

\noindent As $q\to\infty$, the Poissonian weights effectively concentrate around $s=\alpha q\pm\mathcal{O}(\sqrt{\alpha q})$ in \eqref{eq:Large c pi} and $s=\alpha^{-1} q\pm\mathcal{O}(\sqrt{\alpha^{-1} q})$ in \eqref{eq:Large c rho}. Thus, the quantities that appear in the $\delta$-functions in Eqs.~(\ref{eq:Large c pi}) and (\ref{eq:Large c rho}),

\begin{align}
    \bar{\omega}&:=\lambda-\frac{1}{(\alpha q)^2}\sum_{\ell=1}^{s-1} \frac{\alpha^2\tilde{K}_\ell^2}{\sigma_\ell}\ ,
    \label{eq:omegaBar_Def} \\
    \bar{\sigma}&:=1 - \frac{1}{(\alpha^{-1} q)^2}\sum_{\ell=1}^{s-1} \frac{\alpha^{-2}\tilde{K}_\ell^2}{\omega_\ell}\ ,  
    \label{eq:sigmaBar_Def} \\
    \bar{h} &:= \frac{1}{\alpha q}\sum_{\ell=1}^{s-1} \frac{\alpha\tilde{K}_\ell \mu_\ell}{\sigma_\ell}\ , \label{eq:hGaussian} \\
    \bar{\mu} &:=\frac{1}{\alpha^{-1} q}\sum_{\ell=1}^{s-1} \frac{\alpha^{-1}\tilde{K}_\ell h_\ell}{\omega_\ell} \label{eq:muGaussian}\ ,
\end{align} 

\noindent are non-fluctuating in the limit, due to the law of large numbers. Consequently, the $\delta$-functions force the densities to concentrate around $(\omega,h)=(\bar{\omega},\bar{h})$ and $(\sigma,\mu)=(\bar{\sigma},\bar{\mu})$,

\begin{align}
    \pi(\omega,h) &= \delta(\omega-\bar{\omega}) \delta(h-\bar{h})\ , \label{eq:pi conc}\\
    \rho(\sigma,\mu) &= \delta(\sigma-\bar{\sigma}) \delta(\mu-\bar{\mu})\label{eq:rho conc}\ .
\end{align}

\noindent This fact, in turn, enables us to evaluate $\bar{\omega},\bar{\sigma},\bar{h}$ and $\bar{\mu}$ self-consistently, by substituting $\omega_\ell =\bar{\omega}$, $\sigma_\ell =\bar{\sigma}$, $h_\ell=\bar{h}$ and $\mu_\ell = \bar{\mu}$ into Eqs. \eqref{eq:omegaBar_Def} - \eqref{eq:muGaussian},

\begin{align}
    \bar{\omega} &= \lambda + \mathcal{O}(q^{-1})
    \label{eq:OmegaBarEq}\ , \\
    \bar{\sigma} &= 1 + \mathcal{O}(q^{-1})
    \label{eq:SigmaBarEq}\ , \\
    \bar{h} &= \frac{\alpha\langle \tilde{K}\rangle\bar{\mu}}{\bar{\omega}}\ ,
    \label{eq:hBarEq} \\
    \bar{\mu} &= \frac{\alpha^{-1}\langle \tilde{K}\rangle\bar{h}}{\bar{\sigma}}\ .
    \label{eq:muBarEq}
\end{align}
At this point, it becomes clear why the two cases—central and noncentral—require different scalings of the nonzero entries in order to obtain $\langle \lambda_1 \rangle = \mathcal{O}(1)$. In the noncentral case, where $\langle K \rangle \ne 0$, the moments of the variables $h$ and $\mu$ would diverge in the $q\to\infty$ limit if we were to choose $K = \tilde{K}/\sqrt{q}$. This divergence would, in turn, lead to a diverging $\langle \lambda_1 \rangle$. On the contrary, if $\langle K\rangle = 0$, the scaling $K=\tilde{K}/q$ would lead to a vanishing moments of the variables $h$ and $\mu$ in the $q\to\infty$ limit, hence to a vanishing $\langle \lambda_1 \rangle$.

Solving Eqs. \eqref{eq:OmegaBarEq} - \eqref{eq:muBarEq} for $\bar{\omega},\bar{\sigma}$ and $\lambda$ we obtain

\begin{align}
    \bar{\omega} &= \langle \tilde{K} \rangle^2 + \mathcal{O}(q^{-1})\ , \label{eq:OmegaBar sol}\\
    \bar{\sigma} &= 1 + \mathcal{O}(q^{-1})\ , \label{eq:SigmaBar sol}\\
    \lambda &= \langle \tilde{K} \rangle^2 + \mathcal{O}(q^{-1})\ .
\end{align}

\noindent Recalling that $\langle \lambda_1\rangle = \lambda$, we finally get

\begin{equation}
    \langle \lambda_1\rangle =  \langle \tilde{K} \rangle^2 + \mathcal{O}(q^{-1})\ ,
\end{equation}

\noindent which coincides with the isolated largest eigenvalue of the noncentral Wishart ensemble (see Eq. (52) in Ref. \cite{Vinayak2014}, with the appropriate scaling).

To obtain the density of the top eigenvector's components in the dense limit, we start from Eq. \eqref{eq:density_vector_general1}. After rescaling the weights as $K_{ij}=\tilde{K}_{ij}/q$ and accounting for the fact that $\rho(\sigma,\mu)$ and $\pi(\omega,h)$ concentrate [see Eqs. \eqref{eq:rho conc} and \eqref{eq:pi conc}], we obtain

\begin{equation}
T(u)  =\sum_{s=0}^{\infty}p_{\alpha q}(s)\left\langle\delta\left(u- \frac{\alpha\bar{\mu}}{\bar{\omega}\bar{\sigma}}\frac{1}{\alpha q}\sum_{\ell=1}^s\tilde{K}_\ell \right)\right\rangle_{\{K\}_s}\ .\label{eq:T(u) rescaled}
\end{equation}
Again, as $q\to\infty$, the Poissonian weights concentrate around $s=\alpha q\pm\mathcal{O}(\sqrt{\alpha q})$. Hence, the quantity

\begin{equation}
    \bar{u}= \frac{\alpha\bar{\mu}}{\bar{\omega}\bar{\sigma}}\frac{1}{\alpha q}\sum_{\ell=1}^s\tilde{K}_\ell
    \label{eq:u Dense}
\end{equation}
is again non-fluctuating, due to the law of large numbers. Consequently, the $\delta$ function in Eq. \eqref{eq:T(u) rescaled} forces $T(u)$ to concentrate around $u=\bar{u}$, which evaluates to

\begin{equation}
    \bar{u}=\frac{\alpha\bar{\mu}}{\bar{\sigma}\bar{\omega}}\langle \tilde{K} \rangle\ .
    \label{eq:u Mean Dense}
\end{equation}
Combining the concentration of the top eigenvector's components with the normalisation of the eigenvectors, $|\bm{v}|^2 = M$, we expect that $\bar{u} = 1$. This result indeed follows directly from evaluating $\bar{\mu}$. Since the first two equations of \eqref{eq:system} determine the distribution of $\mu$'s up to an arbitrary scaling, to fix the value of $\bar{\mu}$ we use the integral normalisation condition [third line in Eq. \eqref{eq:system}]. After rescaling the weights and accounting for the fact that $\pi(\sigma,\mu)$ and $\rho(\omega,h)$ concentrate, it takes the form 

\begin{equation}
    \sum_{s=0}^\infty p_{\alpha q}(s)\frac{\alpha^2\bar{\mu}^2}{\bar{\omega}^2\bar{\sigma}^2} \Big\langle \frac{1}{(\alpha q)^2}\left( \sum_{\ell=1}^s \tilde{K}_\ell \right)^2\Big\rangle_{\{K\}_s} = 1\ .
    \label{eq:Norm dense}
\end{equation}
Evaluating the average over the weights, we obtain

\begin{equation}
        \frac{\alpha^2\bar{\mu}^2}{\bar{\omega}^2\bar{\sigma}^2}\sum_{s=0}^\infty p_{\alpha q}(s)  \frac{s(s-1)\langle \tilde{K}\rangle^2+s\langle \tilde{K}^2\rangle}{(\alpha q)^2}= 1\ .
\end{equation}
Using the known moments of the Poisson distribution, we have
\begin{equation}
   \frac{\alpha^2\bar{\mu}^2}{\bar{\sigma}^2\bar{\omega}^2}\langle \tilde{K}\rangle^2 + \mathcal{O}\left( q^{-1} \right) = 1\ .
    \label{eq:Dense 1}
\end{equation}
Substituting Eq. \eqref{eq:Dense 1} into \eqref{eq:u Mean Dense}, we obtain $\bar{u}=1$ as anticipated, such that

\begin{equation}
    T(u)=\delta(u-1)\ .
    \label{eq:T(u) FINAL}
\end{equation}
Hence, in the \emph{dense limit}\footnote{{Note that this corresponds to a sequence of two limits: first, $N,M\to\infty$ (with their ratio fixed), and next $q\to\infty$.}} of our noncentral model, the top eigenvector is fully localised around $\bm{1}=(1,...,1)\in \mathbb{R}^M$. Incidentally, we note that this localisation phenomenon is very similar to Ref.~\cite{Lang1964}, valid for the slightly different setting of dense symmetric random matrices with \textit{independent} entries drawn from an arbitrary distribution. 

For \emph{finite} $N,M$, we could have defined the density of top eigenvector's components as $T_{N,M} \left(u \mid q, p(K) \right)$. Our previous result would then correspond to computing the double limit
\begin{equation}
\lim_{q\to\infty}\lim_{M\to\infty} T_{\alpha^2 M,M} \left(u \mid q, p(K) \right)=\delta(u-1)\ . 
\end{equation}

\begin{figure}
    \centering
    \includegraphics[width=0.45\linewidth]{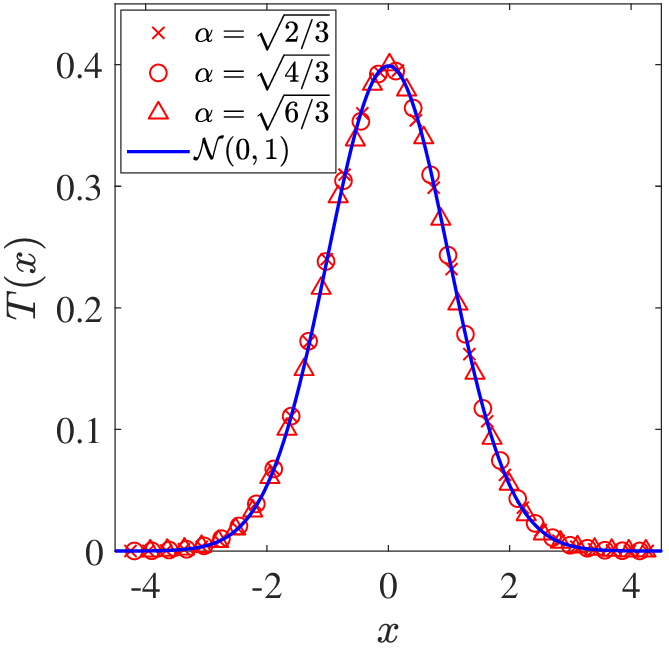}
    \caption{Density of the top eigenvector's components in the dense regime (i.e., $q = \alpha M$), plotted as a function of the rescaled variable $x = \delta\sqrt{M}(u - 1)$, where $\delta$ is defined in Eq.~\eqref{eq: delta Var}. The numerical data (symbols) were obtained by diagonalizing 20 independent realizations of the matrix $\bm{J} = \bm{X}^T \bm{X}$, where $\bm{X}$'s entries are drawn from a uniform distribution $p(K) = \Theta(K)\Theta(1 - K)$. We fixed $M = 4 \cdot 10^4$ and examined three values of $N$, corresponding to ($\times$) $\alpha = \sqrt{2/3}$, ($\circ$) $\alpha = \sqrt{4/3}$, and ($\triangle$) $\alpha = \sqrt{6/3}$. The numerical results exhibit excellent agreement with the standard normal distribution $\mathcal{N}(0,1)$ (solid line).}
    \label{fig:Gaussian fluctuations}
\end{figure}

Empirically, numerical diagonalisation on large but finite matrices shows that $T(u)$ indeed becomes narrower as $q$ is gradually increased. However, it would be interesting to study how this complete localisation in the limit is approached on a narrower scale as $N,M$ increase. While the finite $N,M$ density $T_{N,M} \left(u \mid q, p(K) \right)$ is not attainable via our method, we nevertheless conjecture (on the basis of numerical simulations) that the components of the top eigenvector display Gaussian fluctuations in the double-scaling limit
\begin{equation}
T(x):=\lim_{M\to\infty}\frac{1}{\delta\sqrt{M}} T_{\alpha^2 M,M}\left(u = 1 +\frac{1}{\sqrt{M}}\frac{x}{\delta}\mid q = \alpha M,p(K)\right)=\mathcal{N}(0,1)
\end{equation}
with
\begin{equation}
    \delta=\frac{\alpha}{\sqrt{\frac{\langle\tilde{K}^2\rangle}{\langle\tilde{K}\rangle^2}-1}}\ .
    \label{eq: delta Var}
\end{equation}
In Fig. \ref{fig:Gaussian fluctuations}, we numerically validate our conjecture. The figure shows the density of the top eigenvector's components in the dense regime (i.e., $q = \alpha M$), plotted as a function of the rescaled variable $x = \delta\sqrt{M}(u - 1)$, where $\delta$ is defined in Eq.~\eqref{eq: delta Var}. The numerical data (symbols) were obtained by diagonalizing 20 independent realisations of the matrix $\bm{J} = \bm{X}^T \bm{X}$, where $\bm{X}$'s entries are  drawn from a uniform distribution $p(K) = \Theta(K)\Theta(1 - K)$. We fixed $M = 4 \cdot 10^4$ and examined three values of $N$, corresponding to ($\times$) $\alpha = \sqrt{2/3}$, ($\circ$) $\alpha = \sqrt{4/3}$, and ($\triangle$) $\alpha = \sqrt{6/3}$. The numerical results exhibit excellent agreement with the standard normal distribution $\mathcal{N}(0,1)$ (solid line). A first-principles proof of this conjecture would be very welcome.

\color{black}
\section{Summary and conclusions}
\label{Conclusions}

In summary, we developed a replica formalism to compute the top eigenpair statistics of sparse correlation matrices of the form \( \bm{X}^T \bm{X} \), where the nonzero entries follow a nonzero-mean weight distribution \( p(K) \), leading to an isolated largest eigenvalue.  

Specifically, we focused on the average largest eigenvalue and the density of its associated eigenvector components. The problem of evaluating the average largest eigenvalue can be reformulated as an optimisation problem involving a quadratic Hamiltonian on the sphere. In the zero-temperature limit \( \beta \to \infty \), the Gibbs measure concentrates around the ground state, corresponding to the top eigenvector. Using the replica method, we evaluated the disorder-averaged partition function and derived a system of self-consistent equations governing the order parameter \( \lambda \) [see Eq.~\eqref{eq:system}].  

We solved these equations via a population dynamics algorithm and identified \( \langle \lambda_1 \rangle \), the average largest eigenvalue, as the critical value of \( \lambda \) that determines the convergence of the population dynamics: for \( \lambda < \langle \lambda_1 \rangle \), variables diverge, while for \( \lambda > \langle \lambda_1 \rangle \), they converge to zero. Numerical simulations confirmed excellent agreement between this critical value and direct numerical diagonalisation, both for the degenerate case \( p(K) = \delta_{K,1} \) and for a uniform weight distribution over \( K \in [0,1] \).  

Building on this, we extended our method to compute the density of top eigenvector components. Again, numerical results showed excellent agreement with diagonalisation for both weight distributions.  

Finally, we demonstrated that taking the appropriate dense limit of our model recovers known results from the noncentral Wishart ensemble. Future work should explore the non-gapped regime and the connection between \( p(K) \) and the detachment transition.

\color{black}
\newpage

\section*{Acknowledgment}
P.V. acknowledges support from UKRI FLF Scheme (No. MR/X023028/1).

\appendix
\section{Upper Bound for $\Big\langle\lambda_1\Big\rangle$}
\label{Upper Bound}
In this appendix we show that 

\begin{equation}
    \langle \lambda_1 \rangle\leq \left[\max \left(|\zeta^-|,|\zeta^+|\right)\right]^2 RC\ ,
    \label{eq:Upper bound}
\end{equation}

\noindent where $\zeta^-$ ($\zeta^+$) is the lower (upper) bound of the support of $p(K)$, while $R$ and $C$ are the maximal numbers of nonzero elements in each row and in each column respectively. \color{black} Our starting point is the identification of $\lambda_1$ with the square of the spectral norm of the matrix $\bm{X}$. According to identity 15.511.1 from \cite{gradshteyn2007}, the spectral norm obeys

\begin{equation}
    \lambda_1 \leq\left( \max_{1\leq j\leq M} \sum_{i=1}^N |X_{ij}|\right) \left(\max_{1\leq i\leq N} \sum_{j=1}^M |X_{ij}|\right)\ .
\end{equation}

\noindent Since $X_{ij}=c_{ij}K_{ij}$ [Eq. \eqref{eq:X Def}], we can use the fact that $p(K)$ has a bounded support and that the number of nonzero elements in each row (column) is restricted by $R$ ($C$) to write 

\begin{align}
    \lambda_1 &\leq \left[\max \left(|\zeta^-|,|\zeta^+|\right)\right]^2 \left( \max_{1\leq j\leq M} \sum_{i=1}^N c_{ij}\right) \left(\max_{1\leq i\leq N} \sum_{j=1}^M c_{ij}\right) \nonumber \\
    &\leq \left[\max \left(|\zeta^-|,|\zeta^+|\right)\right]^2 RC \ . 
    \label{eq:UpperBoundSums}
\end{align}

 \noindent Since this inequality holds for every realisation of $\bm{X}$, the ensemble average of $\lambda_1$ clearly satisfies this condition too. Hence, we obtained our desired result, Eq. \eqref{eq:Upper bound}. 
\color{black}

\section{Performing the Average in \eqref{eq:J Average}}
\label{Average}
In this appendix, we show how to compute the average 

\begin{equation}
     \Big\langle \prod_{i=1}^M \prod_{j=1}^N \exp\left(\beta X_{ji}\sum_{a=1}^n v_{ia}u_{ja}\right)\Big\rangle\ ,
\end{equation}

\noindent in the $q\ll \sqrt{NM}$ limit. This average is performed over different realisations of the $N\times M$ random matrix $\bm{X}$, whose entries are i.i.d random variables, expressed as $X_{ji}=c_{ji}K_{ji}$, and are drawn from 

\begin{equation}
    P\left( X_{ji} \right) = \left[\frac{q}{\sqrt{NM}}\delta_{c_{ji},1}+\left(1-\frac{q}{\sqrt{NM}}\right)\delta_{c_{ji},0}\right]p\left(K_{ji}\right)\ ,
\end{equation}
\noindent with $p(K)$ being the weight distribution.  First, we use the independence of the entries to factorise the average,

\begin{equation}
     \Big\langle \prod_{i=1}^M \prod_{j=1}^N \exp\left(\beta X_{ji}\sum_{a=1}^n v_{ia}u_{ja}\right)\Big\rangle = \prod_{i=1}^M \prod_{j=1}^N  \Big\langle\exp\left(\beta cK\sum_{a=1}^n v_{ia}u_{ja}\right)\Big\rangle_{c,K},
\end{equation}
\noindent where $\Big\langle\cdot\Big\rangle_{c,K}$ denotes averaging over a single instance of the random variables $c$ and $K$. Next, we average over the $c$'s and take the $q\ll\sqrt{NM}$ limit to obtain

\begin{align}
\Big\langle \prod_{i=1}^M \prod_{j=1}^N \exp\left(\beta          
 X_{ji}\sum_{a=1}^n v_{ia}u_{ja}\right)\Big\rangle &=  \prod_{i=1}^M \prod_{j=1}^N \left[1 + \frac{q}{\sqrt{NM}}\left(\Big\langle \mathrm{e}^{\beta K\sum_{a=1}^n v_{ia}u_{ja}}\Big\rangle_K -1 \right) \right] \nonumber \\
 & \simeq \exp \left[ \frac{q}{\sqrt{NM}}\sum_{i=1}^M\sum_{j=1}^N \left(\Big\langle \mathrm{e}^{\beta K\sum_{a=1}^n v_{ia}u_{ja}}\Big\rangle_K -1 \right) \right],
\end{align}

\noindent which matches the result in \eqref{eq:J Average}.

\end{document}